\documentclass[journal=jacsat,manuscript=article]{achemso}

\usepackage{graphicx}
\usepackage{xcolor}
\usepackage{float}
\usepackage[utf8]{inputenc}
\usepackage{babel}
\usepackage[T1]{fontenc}
\usepackage{tabularx,booktabs,multirow}
\usepackage{amsmath,amssymb,mathtools,bm}
\usepackage{siunitx}	
\usepackage{mciteplus}
\usepackage{pdfpages}

\newcommand{\reftextit}[1]{}

\newcommand{\ket}[1]{\left| #1 \right>}

\newcommand{\bk}{\mathbf{k}}

\newcommand{\bq}{\mathbf{q}}

\newcommand{\bR}{\mathbf{R}}

\author{Niklas Hofmann}
\affiliation{Institute for Experimental and Applied Physics, University of Regensburg, 93040 Regensburg, Germany}

\author{Alexander Steinhoff}
\affiliation{Institute for Theoretical Physics, Universit\"at Bremen, P.O. Box 330 440, 28334 Bremen, Germany}
\alsoaffiliation{Bremen Center for Computational Materials Science, Universit\"at Bremen, 28334 Bremen, Germany}

\author{Razvan Krause}
\affiliation{Institute for Experimental and Applied Physics, University of Regensburg, 93040 Regensburg, Germany}
		
\author{Neeraj Mishra}
\affiliation{Center for Nanotechnology Innovation@NEST, Istituto Italiano di Tecnologia, Pisa, Italy}
\alsoaffiliation{Graphene Labs, Istituto Italiano di Tecnologia, Genova, Italy}
	
\author{Giorgio Orlandini}
\affiliation{Center for Nanotechnology Innovation@NEST, Istituto Italiano di Tecnologia, Pisa, Italy}

\author{Stiven Forti}
\affiliation{Center for Nanotechnology Innovation@NEST, Istituto Italiano di Tecnologia, Pisa, Italy}
		
\author{Camilla Coletti}
\affiliation{Center for Nanotechnology Innovation@NEST, Istituto Italiano di Tecnologia, Pisa, Italy}
\alsoaffiliation{Graphene Labs, Istituto Italiano di Tecnologia, Genova, Italy}

\author{Tim O. Wehling}
\affiliation{I. Institute of Theoretical Physics, University of Hamburg, Notkestrasse 9, 22607 Hamburg, Germany}
\alsoaffiliation{The Hamburg Centre for Ultrafast Imaging, Luruper Chaussee 149, 22761 Hamburg, Germany}

\author{Isabella Gierz}
\email{isabella.gierz@ur.de}
\affiliation{Institute for Experimental and Applied Physics, University of Regensburg, 93040 Regensburg, Germany}

\title{k-resolved ultrafast light-induced band renormalization in monolayer WS$_2$ on graphene}

\abbreviations{trARPES, DFT, GW}

\begin{document}

\pagebreak

\begin{abstract}

Understanding and controlling the electronic properties of two-dimensional materials is crucial for their potential applications in nano- and optoelectronics. Mono\-layer transition metal dichalcogenides such as WS$_2$ have garnered significant interest due to their strong light-matter interaction and extreme sensitivity of the band structure to the presence of photogenerated electron-hole pairs. In this study, we investigate the transient electronic structure of monolayer WS$_2$ on a graphene substrate after resonant excitation of the A-exciton using time- and angle-resolved photoemission spectroscopy. We observe a pronounced band structure renormalization including a substantial reduction of the transient band gap that is in good quantitative agreement with our {\it ab initio} theory that reveals the importance of both intrinsic WS$_2$ and extrinsic substrate contributions to the transient band structure of monolayer WS$_2$. Our findings not only deepen the fundamental understanding of band structure dynamics in two-dimensional materials but also offer valuable insights for the develop\-ment of novel electronic and optoelectronic devices based on monolayer TMDs and their hetero\-structures with graphene.

\end{abstract}
	
\pagebreak

	
	The size of the band gap of a semiconductor is the essential parameter that determines its optical and electronic properties with crucial importance for light absorption in solar cells and photodetectors as well as the performance of transistors and other electronic devices. In two-dimensional (2D) semiconductors such as monolayer transition metal dichalcogenides (TMDs), confinement and reduced screening give rise to strong electronic correlations that modify the size of the quasiparticle gap and lead to the formation of excitons with large binding energies that dominate the optical response. Controlling the band gap of 2D semiconductors is highly relevant for the development of next generation optoelectronic devices. This is commonly done by tailoring the dielectric screening. At equilibrium, this has been achieved by embedding 2D semiconductors in van der Waals (vdW) heterostructures \cite{Ugeda2014,Waldecker2019} or by changing the carrier density with a gate voltage \cite{Qiu2019}. Out of equilibrium, photodoping with femtosecond laser pulses has been used to modify the screening dynamically, which offers the possibility to control the size of the band gap on femtosecond time scales \cite{Chernikov2015,Pogna2016,Ulstrup2016,Cunningham2017,Liu2019,Krause2021,Lin2022,Dong2023,Hofmann2023}.
	
	{\it Ab initio} theory predicts that the band shifts caused by the presence of excited carriers are non-rigid with a pronounced momentum dependence \cite{Erben2018,Erben2022}. To date, however, experiments have mainly probed the transient size of the direct gap at the K-point. A possible momentum dependence of the transient gap remains unexplored.
	
	Here, we use time- and angle-resolved photoemission spectroscopy (trARPES) to probe the transient band structure of a 2D semiconductor over the whole Brillouin zone. For this purpose, we use monolayer WS$_2$ on a graphene/SiC(0001) substrate that we excite at resonance to the A-exciton at $\hbar\omega_{pump}=2$ eV. We extract the size of the direct quasiparticle gap at the K-point as well as the momentum-resolved valence band (VB) shift as a function of time. The direct gap is found to shrink by $\Delta E\sim 140$\,meV for a fluence of $1.5$\,mJ\,cm$^{-2}$. Within the experimental error bars, the transient VB shift is found to be rigid between $\Gamma$ and $\mathrm{K}$ with a fluence-dependent amplitude ranging from $\Delta E\sim 100$\,meV to $\Delta E\sim 170$\,meV for fluences between $0.7$\,mJ\,cm$^{-2}$ and $1.5$\,mJ\,cm$^{-2}$. Our trARPES results further provide access to the transient carrier distribution and temperature that serve as input parameters for advanced {\it ab initio} calculations that include Hartree and GW contributions of WS$_2$ as well as GdW contributions of the graphene substrate. The calculations quantitatively reproduce the experimentally observed shifts. In addition, our theory allows us to disentangle the relevance of the individual contributions and to determine the momentum dependence of the band shifts.
	
	The detailed microscopic understanding gained in this work provides important information for the design of next generation optoelectronic devices.

	
	\textbf{Sample growth}: 4H-SiC substrates were H-etched to remove scratches and subsequently graphitized in Ar atmosphere. The resulting carbon monolayer with ($6\sqrt{3}\times6\sqrt{3}$)\,R\,$30$$^{\circ}$ structure was decoupled from the SiC substrate by H-intercalation, yielding quasi-freestanding monolayer graphene on H-terminated SiC(0001) \cite{Riedl2009}. WS$_2$ was then grown by chemical vapor deposition from solid WO$_3$ and S precursors \cite{Forti2017}. Atomic force microscopy (AFM) and secondary electron microscopy (SEM) revealed that WS$_2$ grows in the shape of triangular islands with side lengths in the range of $300-700$\,nm with well-defined twist angles of either $0$$^{\circ}$ or $30$$^{\circ}$ with respect to the graphene layer \cite{Hofmann2023}.
	
	\textbf{trARPES}: The setup was based on a commercial titanium sapphire amplifier (Astrella, Coherent) with a central wavelength of $800$\,nm, a repetition rate of $1$\,kHz, a pulse duration of $35$\,fs, and a pulse energy of $7$\,mJ. $5$\,mJ were used to seed a commercial optical parametric amplifier (Topas Twins, Light Conversion) the signal output of which was frequency doubled, yielding $2$\,eV pump pulses resonant with the A-exciton of monolayer WS$_2$. The remaining $2$\,mJ of output energy were frequency doubled and focused onto an argon gas jet for high harmonic generation. A single harmonic at $21.7$\,eV photon energy was selected with a grating monochromator yielding extreme ultraviolet (XUV) probe pulses that were used to eject photoelectrons from the sample. The photoelectrons were dispersed according to their kinetic energy and emission angle by a hemispherical analyzer (Phoibos 100, SPECS), yielding two-dimensional snapshots of the occupied part of the band structure in momentum space. The probe spot diameter was $\sim250$\,$\mu$m on the sample, covering many different WS$_2$ islands. Nevertheless, $0$$^{\circ}$ and $30$$^{\circ}$ WS$_2$ islands were easily distinguished based on the dispersion of their band structure in momentum space. The energy and temporal resolutions for the measurements presented in the present publication were $\sim200$\,meV and $160$\,fs, respectively.
	
	\textbf{Theory}: We combine non-equilibrium Green functions with {\it ab initio} calculations of the ground state properties to compute the influence of photoexcited electron-hole pairs on the transient electronic structure of monolayer WS$_2$ on a graphene/SiC substrate across the whole Brillouin zone. Electrons and holes are assumed to follow a quasi-thermal distribution with one common elevated temperature. The influence of excited carriers inside the WS$_2$ layer is treated explicitely in the GW self-energy. The contribution of excited carriers inside the graphene layer enters via a macroscopic dielectric function. We consider both Hartree and static as well as dynamical exchange renormalizations. Further details are provided in the Supporting Information (SI).

\begin{figure}
		\includegraphics[width=\columnwidth]{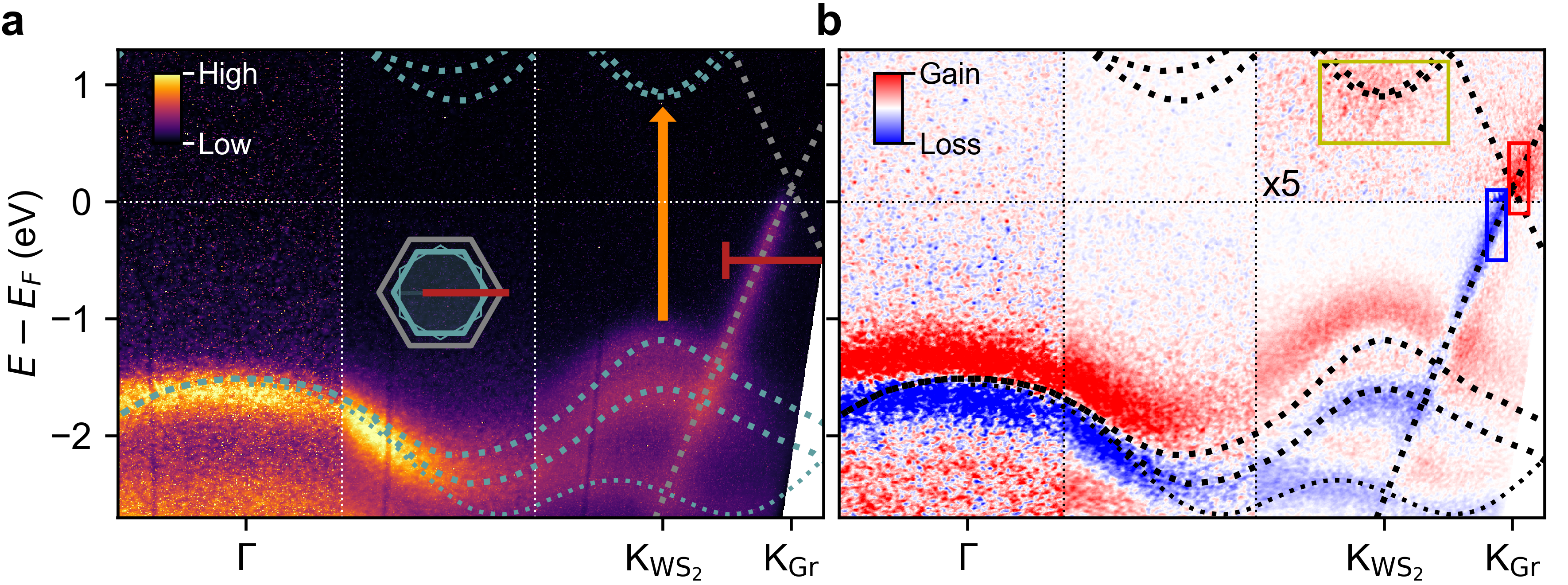}
		\caption{\textbf{trARPES data of WS$_2$-graphene heterostructure.} \textbf{a)} ARPES spectrum measured at negative pump-probe delay before the arrival of the pump pulse along the $\Gamma\mathrm{K}$ direction as indicated by the red line in the inset. The orange vertical arrow illustrates the excitation resonant to the A-exciton at $\hbar\omega_\text{pump}= 2.0$\,eV. Gray and green dashed lines indicate the theory band structures for graphene \cite{Wallace1947} and WS$_2$ \cite{Zeng2013}, respectively, that were shifted in energy to match the observed band alignment. The thin dashed green line marks the band structure of WS$_2$ flakes with 30$^{\circ}$ rotation relative to the graphene layer. The horizontal red line covers the $k$ range over which EDCs in Fig. 3b) are extracted. \textbf{b)} Pump-induced changes $250$ fs after excitation at $\hbar\omega_\text{pump}=2.0$\,eV with a fluence of 1.5\,mJ\,cm$^2$. Red and blue indicate a gain and loss of photocurrent with respect to negative pump-probe delays, respectively. The upper right panel is multiplied with a factor of 5 for better visibility. Colored boxes indicate the area of integration for the pump-probe traces in Fig. 3a).}
		\label{figure1}
	\end{figure}

	
	Figure \ref{figure1}a shows the band structure of the WS$_2$-graphene sample measured along the $\Gamma\mathrm{K}$-direction of the $0$$^{\circ}$ WS$_2$ islands at negative pump-probe delay before the arrival of the pump pulse. Gray and green dashed lines are theoretical band structures from \cite{Zeng2013} for monolayer WS$_2$ and from \cite{Wallace1947} for monolayer graphene that have been shifted in energy to account for the experimentally observed doping and equilibrium gap size. The thin dashed green line indicates the dispersion of the $30$$^{\circ}$ WS$_2$ islands. The orange arrow highlights the direct electronic transition triggered by photoexcitation at $\hbar\omega_\text{pump}=2$ eV. Figure \ref{figure1}b shows the pump-induced changes of the photocurrent at a pump-probe delay of $t=250$\,fs for a pump fluence of $F=1.5$\,mJ\,cm$^{-2}$. Red and blue indicate gain and loss, respectively, with respect to the unperturbed photocurrent in Fig. \ref{figure1}a. The pump-probe signal exhibits three main features: (1) The minimum of the WS$_2$ conduction band (CB) at $\mathrm{K}_\text{WS\textsubscript{2}}$ gets populated by the pump pulse. (2) The WS$_2$ VB exhibits a loss at its equilibrium position and a gain above, indicative of a transient up-shift. Note that the loss at the equilibrium position of the upper valence band is largely compensated by the upshift of the lower valence band. (3) The Dirac cone of graphene shows a loss of photoelectrons below and a gain of photoelectrons above the Fermi level, suggestive of a hot Fermi-Dirac distribution.
	
\begin{figure}
		\includegraphics[width=\columnwidth]{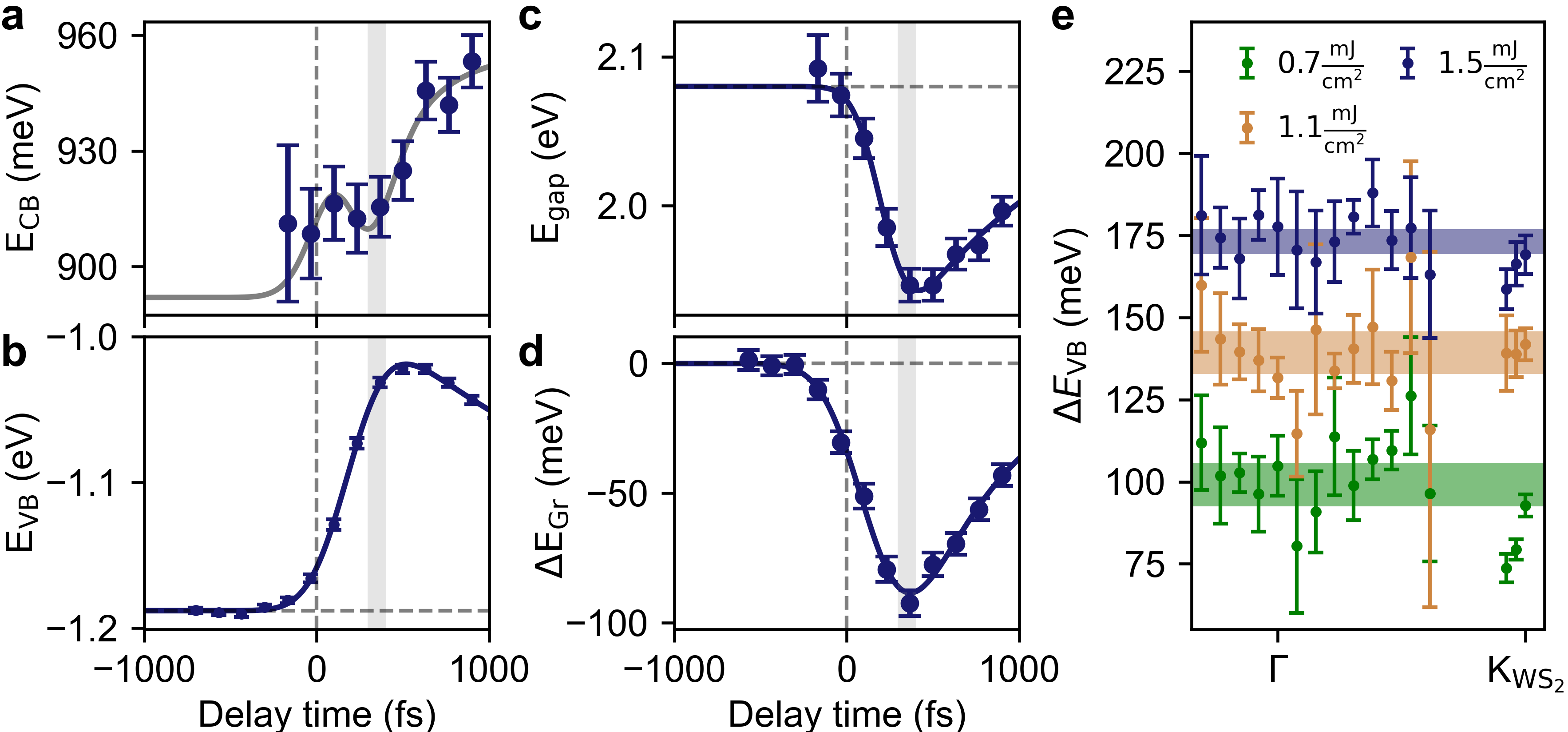}
		\caption{\textbf{trARPES data analysis.} \textbf{a)} Transient CB position. The gray line is a guide to the eye obtained by adding the transient band gap fit from c) to the fit of the VB position from b). \textbf{b)} Transient band shift of the upper VB at the $\mathrm{K}$ point together with exponential decay fit. \textbf{c)} Transient band gap obtained by subtracting the data from a) and b) together with exponential decay fit. \textbf{d)} Graphene band shift together with exponential decay fit. Vertical grey lines in a)-d) mark the pump-probe delay where the VB shift at K reaches its maximum. \textbf{e)} $k$-dependent VB shifts for different excitation fluences for the pump-probe delay where the VB shift at $\mathrm{K}$ reaches its maximum. Horizontal lines represent momentum-averaged shifts. The line width reflects the standard deviation. The error bars in panels a, b d, and e represent the standard deviation determined from the data fitting. The error bars in panel c are the sum of the error bars in a and b.}
		\label{figure2}
	\end{figure}

	In order to extract the transient band gap at $\mathrm{K}_\text{WS\textsubscript{2}}$ and the momentum-resolved VB shift, we proceed as follows: we extract energy distribution curves (EDCs) at different momenta that we fit with an appropriate number of Gaussian peaks and a Shirley background to determine the transient binding energy of the WS$_2$ VB and CB and the Dirac cone of graphene. Further details are provided in the SI. The transient peak positions for the WS$_2$ CB and VB at $\mathrm{K}_\text{WS\textsubscript{2}}$ are shown in Figs. \ref{figure2}a and b, respectively. The transient band gap obtained by subtracting the binding energy of the WS$_2$ VB in Fig. \ref{figure2}b from the binding energy of the WS$_2$ CB in Fig. \ref{figure2}a is shown in Figure \ref{figure2}c together with an exponential fit (see SI). The band gap is found to decrease by $\Delta E_\text{gap} = 140\pm 20$\,meV with a lifetime of $\tau=0.9\pm 0.2$\,ps in good agreement with our own previous results \cite{Krause2021} and slightly lower than typical experimental \cite{Chernikov2015,Ulstrup2016,Liu2019} and theoretical values \cite{Steinhoff2014,Liang2015,Gao2017,Meckbach2018} reported in literature for similar samples. Possible reasons for this minor discrepancy might be related to the use of different substrates and the difficulty to estimate the density of photoexcited electron-hole pairs in the experiment (see below). Figure \ref{figure2}d shows the transient binding energy of the Dirac cone together with an exponential fit (see SI). The Dirac cone is found to shift down by $\Delta E = 90\pm 10$ meV with a lifetime of $\tau=0.6\pm 0.1$ ps. The momentum-resolved shift of the WS$_2$ VB is shown in Fig. \ref{figure2}e for a pump-probe delay of $t_\text{max}\sim 500$\,fs, where the VB shift at $\mathrm{K}_\text{WS\textsubscript{2}}$ reaches its maximum, for three different fluences. Within the error bars, the observed VB shift is found to be momentum-independent with amplitudes of $\Delta E = 100\pm 10$\,meV, $\Delta E = 140\pm 10$\,meV, and $\Delta E = 170\pm 10$\,meV for fluences of $F=0.7$\,mJ\,cm$^{-2}$, $F=1.0$\,mJ\,cm$^{-2}$, and $F=1.5$\,mJ\,cm$^{-2}$, respectively. This is similar to the previously observed rigid band shift for monolayer WS$_2$ resting on different substrates \cite{Waldecker2019} and consistent with previous predictions of {\it ab initio} theory for photoexcited samples \cite{Erben2018, Erben2022}. 
	
\begin{figure}
		\includegraphics[width=\columnwidth]{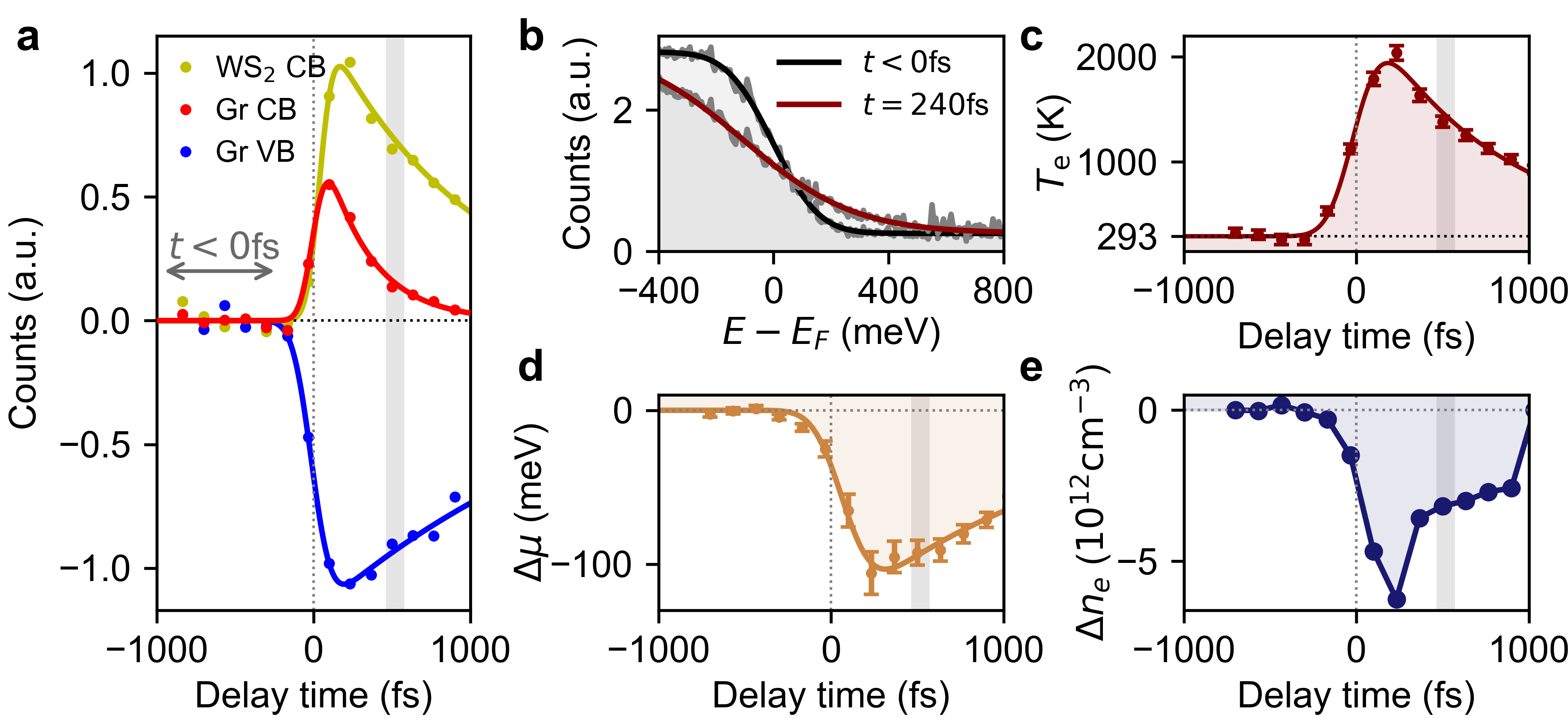}
		\caption{\textbf{Charge transfer dynamics.} \textbf{a)} Photocurrent integrated over the colored boxes from Fig. 1b). The grey arrow marks the integration range for the ARPES spectrum in Fig. 1a). \textbf{b)} Energy distribution curve showing the Fermi edge in the Dirac cone for two different pump-probe delays together with Fermi-Dirac fits. \textbf{c)} Electronic temperature in the Dirac cone as a function of pump-probe delay together with exponential decay fit. \textbf{d)} Chemical potential inside the graphene layer as a function of pump-probe delay together with exponential decay fit. \textbf{e)} Changes in carrier density in the graphene layer, calculated from the electronic temperature from c), the chemical potential from d) and the density of states. Vertical grey lines in a), c), d) and e) mark the pump-probe delay where the VB shift at $\mathrm{K}$ reaches its maximum. The error bars in panels d and e represent the standard deviation determined from the data fitting.}
		\label{figure3}
	\end{figure}

	Next, we determine the non-equilibrium carrier distribution of the WS$_2$-graphene sample at the pump-probe delay that corresponds to the momentum-resolved WS$_2$ VB shift shown in Fig. \ref{figure2}e to provide input for subsequent theory. Figure \ref{figure3}a shows the photocurrent integrated over the three areas marked by colored boxes in Fig. \ref{figure1}b as a function of pump-probe delay together with exponential fits . The Dirac cone of graphene shows a short-lived gain (red, $\tau=300\pm30$\,fs) and a long-lived loss (blue, $\tau=2.10\pm0.03$\,ps). The lifetime of the electrons at the bottom of the WS$_2$ CB (yellow) is found to be $\tau=950\pm70$\,fs. Figure \ref{figure3}b shows the energy-resolved population of the Dirac cone, obtained by integrating the photocurrent over the momentum range indicated by the red scale bar in Fig. \ref{figure1}a, for two different time delays together with Fermi-Dirac fits (see SI). The resulting electronic temperature and chemical potential are shown in Figs. \ref{figure3}c and d, respectively. The electronic temperature reaches a peak value of $T_{e, \text{max}}=1900\pm100$\,K and cools down with an exponential lifetime of $\tau=760\pm60$\,fs. From the electronic temperature and the chemical potential, we calculate the carrier concentration inside the Dirac cone as explained in detail in the SI. The result is shown in Fig. \ref{figure3}e as a function of pump-probe delay. We find that the carrier concentration inside the Dirac cone transiently decreases by $(5.7\pm1.2)\times 10^{12}$\,cm$^{-2}$.
	
	The observations in Figs. \ref{figure2} and \ref{figure3} have been previously attributed to ultrafast charge separation in WS$_2$-graphene heterostructures \cite{Aeschlimann2020,Krause2021,Hofmann2023}. Photoexcitation at resonance to the A-exciton of WS$_2$ is followed by rapid hole transfer into the graphene layer, resulting in a charge-separated transient state with a lifetime of $\sim 1$\,ps. For a pump-probe delay of $t\sim 500$\,fs, where the VB shift at $\mathrm{K}_\text{WS\textsubscript{2}}$ reaches its maximum, we find the following carrier distribution in our WS$_2$-graphene sample: $\sim3.4\times 10^{12}$\,cm$^{-2}$ holes are transferred from WS$_2$ to graphene, while $\sim 70$\% of the photoexcited electrons remain in the WS$_2$ layer. The electrons inside the Dirac cone exhibit an elevated electronic temperature of $T_e\sim 1400$\,K. The values for all fluences investigated in the present work are summarized in Table \ref{table1}.
	
	In Table \ref{table2} we present our estimates for the total density of photo-generated electron-hole pairs $n_{e,h}^{\text{WS}_2}$ for the three pump fluences employed in the experiment. This quantity is difficult to estimate as the absorption of 2D TMDs is highly non-linear due to Pauli blocking and many-body-effects \cite{Erben2022}. According to \cite{Erben2022}, $n_{e,h}^{\text{WS}_2}$ for hBN-encapsulated WS$_2$ is predicted to saturate around $0.5\times 10^{14}$\,cm$^{-2}$ for fluences $\geq0.5$\,mJ\,cm$^{-2}$. A lower limit for $n_{e,h}^{\text{WS}_2}$ is given by the maximum number of holes that are found to be transferred into the graphene layer during ultrafast charge separation.

	\begin{table}[]
		\centering
		\begin{tabularx}{\columnwidth}{|>{\centering\arraybackslash}X|>{\centering\arraybackslash}X|>{\centering\arraybackslash}X|>{\centering\arraybackslash}X|}
			\hline
			pump fluence & $t_\text{max}$ & $n_h^{\mathrm{gr}}$ & $T_e$ \\
			\hline
			0.7\,mJ\,cm$^{-2}$& 450\,fs &$7.9\times 10^{12}$\,cm$^{-2}$ & $1000$\,K\\
			1.0\,mJ\,cm$^{-2}$& 490\,fs & $8.8\times 10^{12}$\,cm$^{-2}$ & $1300$\,K\\
			1.5\,mJ\,cm$^{-2}$& 520\,fs & $10.4\times 10^{12}$\,cm$^{-2}$ & $1400$\,K\\
			\hline
		\end{tabularx}
	\caption{time delay $t_\text{max}$, where WS$_2$ VB shift at $\mathrm{K}$ reaches its maximum; hole density inside Dirac cone $n_h^{gr} (t=t_\text{max})$; electronic temperature of carriers inside Dirac cone $T_e(t=t_\text{max})$.}
	\label{table1}
	\end{table}
	
	\begin{table}[]
		\centering
		\begin{tabularx}{\columnwidth}{|>{\centering\arraybackslash}X|>{\centering\arraybackslash}X|>{\centering\arraybackslash}X|}
			\hline
			pump fluence & $n_{e,h}^{\mathrm{WS}_2}$ upper limit & $n_{e,h}^{\mathrm{WS}_2}$ lower limit \\
			\hline
			0.7\,mJ\,cm$^{-2}$& $5\times 10^{13}$\,cm$^{-2}$ & $2\times 10^{12}$\,cm$^{-2}$ \\
			1.0\,mJ\,cm$^{-2}$& $5\times 10^{13}$\,cm$^{-2}$ & $3\times 10^{12}$\,cm$^{-2}$ \\
			1.5\,mJ\,cm$^{-2}$& $5\times 10^{13}$\,cm$^{-2}$ & $6\times 10^{12}$\,cm$^{-2}$ \\
			\hline
		\end{tabularx}
	\caption{estimated upper and lower bound for the density of photo-generated electron-hole pairs for different fluences.}
	\label{table2}
	\end{table}
	
\begin{figure}[h!t]
    \centering
		\includegraphics[width=\columnwidth]{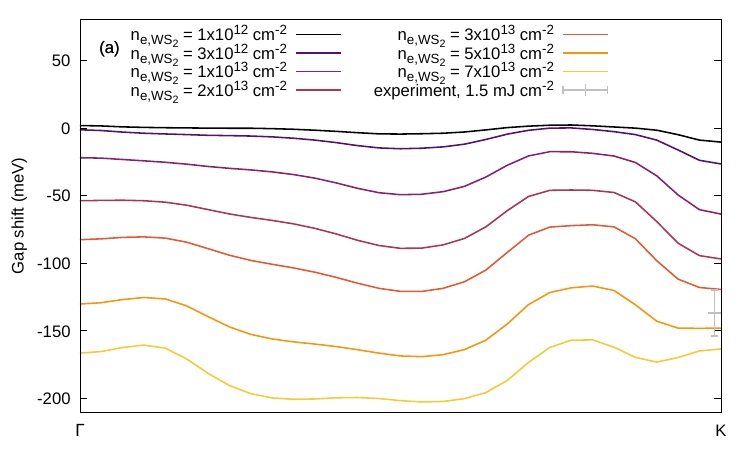}
		\includegraphics[width=\columnwidth]{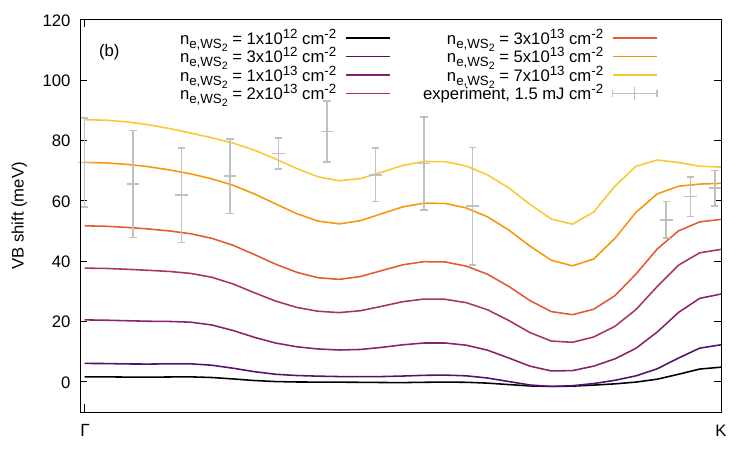}
		\caption{\textbf{Theory.} Calculated k-resolved transient band gap (a) and quasi-particle shifts of WS$_2$ VB (b) for different carrier densities for a carrier temperature of $T=1500$\,K. The grey dots represent charging-shift-corrected experimental data points for a fluence of 1.5\,mJ\,cm$^{-2}$.} 
    \label{figure4}
    \end{figure}

	The parameters in Tables \ref{table1} and \ref{table2} now serve as input for {\it ab initio} calculations of the transient band gap renormalization and WS$_2$ VB shift. At equilibrium, the graphene layer is found to be hole-doped with the Fermi level at $-300$\,meV below the Dirac point (see Fig. \ref{figure1}a) corresponding to a hole concentration of $n_h^{gr,0}=7\times10^{12}$\,cm$^{-2}$. First, we correct the band structure of freestanding monolayer WS$_2$ by adding static GdW corrections due to screening from the graphene/SiC substrate. Next, we compute the transient changes of the WS$_2$ band structure due to screening from the photoexcited electron-hole pairs. For this purpose we assume initial photoexcited electron and hole densities in the range between $n_{e,h}^{\text{WS}_2}=1\times10^{12}$\,cm$^{-2}$ and $n_{e,h}^{\text{WS}_2}=7\times10^{13}$\,cm$^{-2}$ corresponding to the estimates provided in Table \ref{table2}. Further, 90\% of the photoexcited holes are assumed to be transferred into the graphene layer \cite{He2014}. For a WS$_2$ coverage of the graphene/SiC substrate of 50\% \cite{Hofmann2023} this corresponds to a hole density of $n_h^{gr}=n_h^{gr,0}+0.5\times0.9\times n_{e,h}^{\text{WS}_2}$ inside the graphene layer.  Finally, we assume that all carriers have one common electronic temperature in the range between $T=1000$\,K and $T=1400$\,K (see Table \ref{table1}) irrespective of their nature (electron or hole) and their location (WS$_2$ or graphene). Further details about the computational methods are presented in the SI. The results for the size of the transient momentum-resolved 
gap and VB shift at $T=1500$ K are shown in Figs. \ref{figure4}a and b, respectively. We find that the direct WS$_2$ band gap at K$_{\text{WS}_2}$ reduces by $\sim11$\,meV ($\sim163$\,meV) for an electron density of $n_e^{\text{WS}_2}=1\times10^{12}$\,cm$^{-2}$ ($7\times10^{13}$\,cm$^{-2}$) with a k-dependent variation of $\sim13$\,meV ($\sim46$\,meV). The VB shift is found to vary between mean values of $\sim2$\,meV and $\sim70$\,meV in the electron density range from $n_e^{\text{WS}_2}=1\times10^{12}$\,cm$^{-2}$ to $7\times10^{13}$\,cm$^{-2}$ with a k-dependent variation between $\sim6$\,meV and $\sim35$\,meV. Results for $T=1000$ K are shown in the SI in SFig. 7. Note that the calculations do not take into account capacitor-like charging shifts that occur in the transient charge-separated state. For direct comparison between theory and experiment we subtract the charging shift from the experimental data points measured for a fluence of 1.5\,mJ\,cm$^{-2}$ as described in detail in the SI and include them as grey dots in Figs. \ref{figure4}a and b. 


We find that the experimental data points are well reproduced by our calculations for an electron density of $n_e^{\text{WS}_2}=5\times10^{13}$\,cm$^{-2}$ corresponding to the upper limit of the estimated electron-hole pair density. The experimental error bars, however, are too large to allow for an experimental verification of the theoretically predicted k-dependence of the transient band structure changes. In contrast to experiment, our theory allows us to disentangle various different contributions to the transient band structure changes. Based on additional data provided in the SI we conclude that (i) free-standing monolayer WS$_2$ exhibits a VB shift that is roughly constant between $\Gamma$ and $K_{\text{WS}_2}$ but steeply increases at $K_{\text{WS}_2}$ (SFig. 8). This cannot be reconciled with our experimental data in Fig. \ref{figure2}e. (ii) WS$_2$ GW contributions result in a k-dependent WS$_2$ VB shift with three minima along the $\Gamma K$ direction (SFig. 9). The average amplitudes, however, are smaller than observed in experiment. (iii) For quantitative agreement with experiment additional graphene GdW contributions due to ultrafast hole transfer from WS$_2$ to graphene need to be considered (SFig. 10). These yield a WS$_2$ VB shift that is constant for the biggest part of the $\Gamma K$ direction with a minimum at $K_{\text{WS}_2}$ that appears at high carrier densities. We would like to stress that holes located in the WS$_2$ monolayer itself cause much stronger renormalizations than holes located in the relatively remote graphene layer (SFig. 11). (iv) WS$_2$ Hartree contributions are found to be negligible with WS$_2$ VB shifts below 1\,meV (SFig. 12). 
	
	
In summary, we showed that WS$_2$ on graphene exhibits a strong light-induced band structure renormalization that is well reproduced by our {\it ab initio} theory including Hartree and GW contributions of WS$_2$ as well as GdW contributions of the graphene substrate. The experimental error bars, however, are too large to allow for an experimental verification of the theoretically predicted k-dependence of the transient band structure changes. The microscopic insights gained in the present work may guide the development of future opto\-electronic devices based on monolayer TMDs and their heterostructures with graphene.
	
\begin{acknowledgement}
This work received funding from the the European Union's Horizon 2020 research and innovation program under Grant Agreement No. 851280-ERC-2019-STG and 881603-\-Gra\-phene Core3 as well as from the Deutsche Forschungsgemeinschaft (DFG) via the collaborative research center CRC 1277 (project No. 314695032), the Priority Program SPP 2244 (project No. 443405595) and the Research Unit RU 5242 (project No. 449119662). We further acknowledge fruitful discussions with S. Refaely-Abramson as well as resources for compu\-ta\-tional time at the HLRN (G\"ottingen/Berlin).
\end{acknowledgement}

\begin{suppinfo}
The Supporting Information contains details about the data analysis and the theoretical model.
\end{suppinfo}

\clearpage

\clearpage

\section{Supporting Information}

\section{Data Analysis}

\subsection{Extracting the transient band structure}

\begin{figure}[h!t]
	\centering
	\includegraphics[width=\columnwidth]{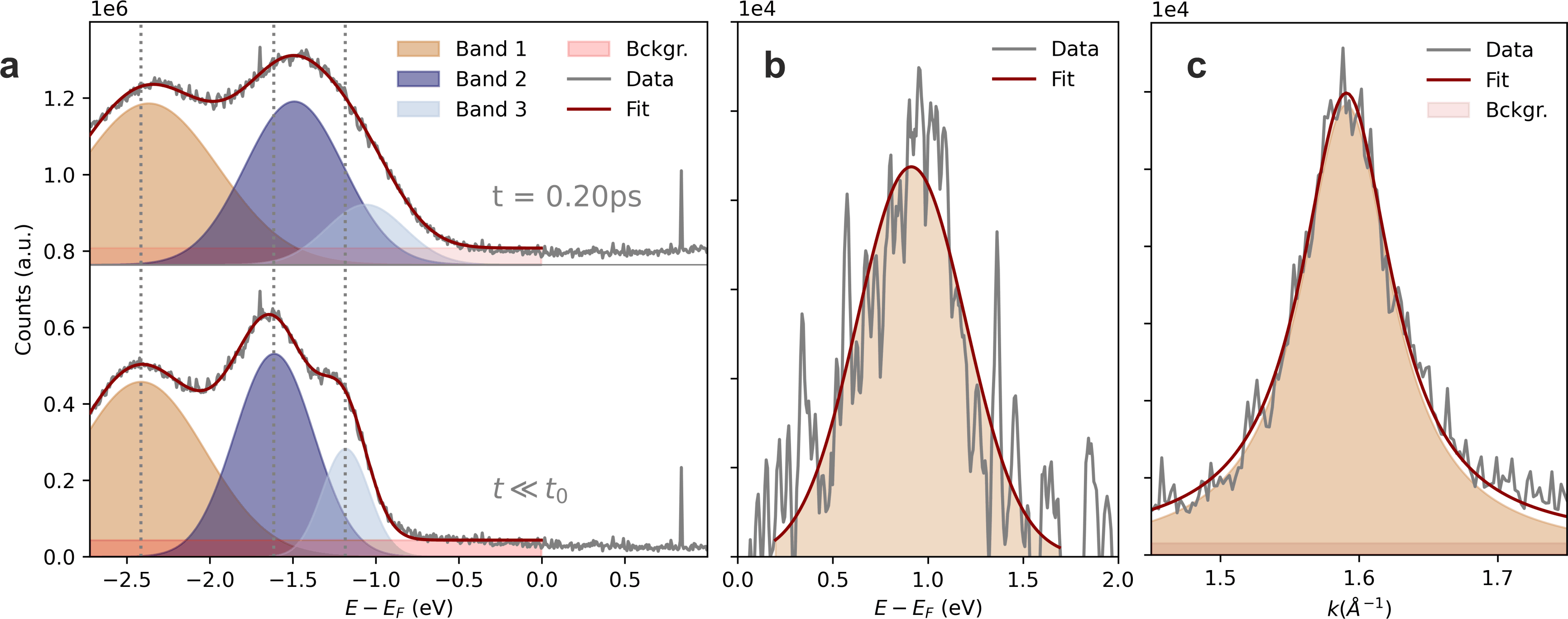}
	\caption{\textbf{Extracting band positions. a)} WS\textsubscript{2} EDCs taken at $k=\SI{1.3}{\per\angstrom}$ for negative pump-probe delay and at $t=\SI{0.2}{\pico\second}$ fitted with a constant offset and three Gaussian peaks. The dotted vertical lines mark the peak positions for negative pump-probe delay. \textbf{b)} Pump-induced changes of the EDC through the WS\textsubscript{2} conduction band together with a Gaussian fit. \textbf{c)} MDC through the graphene Dirac cone at negative pump-probe delay fitted with the sum of a constant offset and a Lorentzian peak.}
	\label{fig:EDCfits}
\end{figure}

In order to extract the transient position of the WS\textsubscript{2} valence band, the data in Fig. 1a (main manuscript) was integrated over the momentum range $\Delta k=\pm\SI{0.04}{\per\angstrom}$ around different momentum positions from $k=\SI{-0.4}{\per\angstrom}$ to $k=\SI{1.3}{\per\angstrom}$. Close to the K-point, the resulting EDCs were then fitted with a constant background and three Gaussian peaks: two for the spin-split WS\textsubscript{2} valence band close to the K-point, and one additional peak for the valence band close to the M-point of the WS\textsubscript{2} flakes with a relative orientation of 30$^{\circ}$ relative to the graphene layer. Exemplary fits are presented in Fig. \ref{fig:EDCfits}a. The following constraints were applied:

\begin{itemize}
	\item The constant background was fixed to the value found for negative pump-probe delay.
	\item The energy difference between the two spin-split WS\textsubscript{2} valence bands was fixed to the value found for negative pump-probe delay.
	\item The spectral weight of the lower spin-split valence band of the flakes with 0$^{\circ}$ orientation and the M-point valence band of the flakes with 30$^{\circ}$ orientation was fixed to the value found for negative pump-probe delay.
\end{itemize}

For momentum positions between $k=\SI{-0.4}{\per\angstrom}$ and $k=\SI{1.0}{\per\angstrom}$, the spin splitting of the valence band is too small to be resolved in our experiment and the valence bands of the WS\textsubscript{2} flakes with both orientations merge together. Thus, the band positions were extracted in the same fashion but using only one Gaussian peak.

To obtain the transient position of the WS\textsubscript{2} conduction band, the data in Fig. 1b (main text) was integrated over the momentum range $\Delta k=\pm\SI{0.05}{\per\angstrom}$ around $k=\SI{1.3}{\per\angstrom}$. The resulting EDCs were then fitted with a Gaussian function (see Fig. \ref{fig:EDCfits}b).

The shift of the graphene Dirac cone was determined by integrating the data in Fig. 1a (main text) over the energy range $\Delta E = \pm\SI{50}{\milli\electronvolt}$ around $(E-E_F )= -\SI{0.6}{\electronvolt}$. The resulting MDCs were fitted by the sum of a constant background and a Lorentzian peak (see Fig. \ref{fig:EDCfits}c). The momentum shift obtained in this way was converted into an energy shift by multiplying with the theoretic slope of the $\pi$-band of \SI{7}{\electronvolt\angstrom}.

In order to subtract the band shift due to layer charging from the valence band shifts for Fig. 4 of the main text, the following procedure was applied for a fluence of $F=\SI{1.5}{\milli\joule\per\centi\meter\squared}$. It was assumed that the band shifts of valence and conduction band due to renormalization is symmetric around the center of the band gap. Then, $E_\text{gap}(t)/2$ was added to the transient valence band position to extract the energy shift of the center of the band gap. This resulted in an upshift of $\num{137}\pm\SI{17}{\milli\electronvolt}$, that has been subtracted from the valence band shifts of Fig. 2 of the main text.

\bigskip

\subsection{Exponential fit of various dynamics}

For extracting rise and decay times of various pump-probe signals in the main text, an analytic fitting function was used. It was obtained by convolving the product of a step function and an exponential decay (describing the underlying dynamics) with a Gaussian function (accounting for the finite temporal resolution):

\begin{equation}
	f(t) = \frac{a}{2} \left( 1 + \mathrm{erf}\left( \frac{(t-t_0)\tau - \frac{\mathrm{FWHM}^2}{8\ln2}}{\sqrt{2}\tau\frac{\mathrm{FWHM}}{2\sqrt{2\ln2}}} \right) \right)
	\exp\left( \frac{\frac{\mathrm{FWHM}^2}{8\ln 2}-2(t-t_0)\tau}{2\tau^2} \right)
\end{equation}

Here, $a$ is the amplitude of the underlying exponential decay, $\mathrm{erf}$ is the error function, $\mathrm{FWHM}$ is the full width at half maximum of the Gaussian describing the width of the rising edge, $t_0$ is the center of the rising edge and $\tau$ is the decay time. This function was used to fit the dynamics showed in Figs. 2b, c and d and Fig. 3 in the main text.

\subsection{Fermi-Dirac fits and transient population of Dirac cone}

For extracting the transient electronic temperature and the transient chemical potential, the time- and angle-resolved photocurrent was integrated over the momentum region specified in Fig. 1 of the main text, yielding the energy-resolved occupation of the Dirac cone. These EDCs were fitted with a Fermi-Dirac distribution convolved with a Gaussian function to account for the finite energy resolution. First, the EDCs at negative delays were fitted with a fixed temperature of \SI{294}{\kelvin} to obtain the width of the Gaussian. Next, the EDCs were fitted with the width of the Gaussian fixed to the value obtained for negative delays, yielding the transient electronic temperature and the transient chemical potential of the graphene layer with respect to the vacuum, $\Delta\mu^\text{vac}$. Subsequently, the energetic shift $\Delta E_\text{gr}$ of the graphene $\pi$-band was obtained as described above (see Fig. \ref{fig:EDCfits}c). The change in chemical potential $\Delta\mu$ relative to the Dirac point is then calculated as $\Delta\mu = \Delta\mu^\text{vac} - \Delta E_\text{gr}$. From there, one can directly extract the change of the total number of electrons in the graphene layer from

\begin{equation}
	\Delta n_e(t) = \int_{-\infty}^{\infty}\rho(E) \left[ f_\text{FD}(E, \mu(t), T(t)) - f_\text{FD}(E, \mu_0, T_0) \right] \,\mathrm{d}E.
\end{equation}

Here, $\rho(E) = \frac{2A_C}{\pi}\frac{\lvert E - E_D \rvert}{\hbar^2 v_F^2}$ is the density of states with the energetic position of the Dirac point $E_D$ and $A_c = \frac{3\sqrt{3}a^2}{2}$ with a lattice constant of $a = \SI{1.42}{\angstrom}$. The number of holes shown in Fig. 3e of the main text is then given by $\Delta n_h(t) = -\Delta n_e(t)$.

\section{Theory}
	
Our theoretical description of atomically thin semiconductors uses the formalism of non-equilibrium Green functions in combination with ab-initio calculations for the ground-state properties. This method provides access to electronic and optical properties of semiconductors under the influence of photoexcited carriers based on material-realistic band structures and interaction matrix elements. Moreover, in a quasi-equilibrium situation frequency-dependent screening effects from a dielectric environment can be systematically taken into account. Hence we do not consider dynamics of charge-carrier populations, but focus on the time window following the optical excitation and relaxation of electrons and holes.

Band structures $\varepsilon_{\bk}^{\lambda}$ of the optically relevant lowest conduction and highest valence bands of WS$_2$ are obtained from a G$_0$W$_0$-calculation as described in Refs.~\cite{steinhoff_influence_2014, steinhoff_exciton_2017}. Here, $\lambda$ denotes the band index, which includes spin for notational simplicity. As an interface between first-principle ground-state and excited-carrier theory we utilize a lattice Hamiltonian formulated in a localized basis of Wannier orbitals $\ket{\alpha}$, where we limit ourselves to the dominant W-d orbitals (d$_{z^2}$, d$_{x^2-y^2}$ and d$_{xy}$). Then, Bloch states are composed of Wannier obitals according to
$\ket{\psi_{\bk}^{\lambda}}=\sum_{\alpha} c^{\lambda}_{\alpha,\bk}\ket{\bk,\alpha} $,
with the coefficients $c^{\lambda}_{\alpha,\bk}$ describing the momentum-dependent contribution of the orbital $\alpha$ to the Bloch band $\lambda$.
The Bloch sums $\ket{\bk,\alpha}$ are connected to the localized basis via $\ket{\bk,\alpha}=\frac{1}{\sqrt{N}}\sum_{\bR}e^{i\bk\cdot\bR}\ket{\bR,\alpha}$ with the number of unit cells $N$ and lattice vectors $\bR$. The background-screened and bare Coulomb interaction can be expressed in matrix form as
$V_{\alpha\beta,\bq}(\omega)$ and $U_{\alpha\beta,\bq}$, respectively.
The valence- and conduction-band splitting caused by spin-orbit interaction is considered along the lines of Ref.~\cite{liu_three-band_2013} and \cite{steinhoff_influence_2014}, including first- and second-order effects.  To take into account dielectric screening by the graphene substrate, we use the \textit{Wannier function continuum electrostatic} (WFCE) approach described in Ref.~\cite{rosner_wannier_2015} that combines a continuum-electrostatic model for the screening with a localized description of Coulomb interaction provided in Ref.~\cite{steinhoff_exciton_2017}. The dynamical screening from free charge carriers in graphene introduces a frequency dependence in the matrix elements of $V$.

\subsection{GW self-energy with frequency-dependent background screening}

On an RPA level, the Schwinger-Keldysh self-energy is \cite{kremp_quantum_2005}
\begin{equation}
\begin{split} 
    \Sigma(1,1')&=\Sigma^{\textrm{H}}(1,1')+\Sigma^{\textrm{GW}}(1,1')\\
     &=-i\hbar\int d2\,V(1,2)G(2,2^+)\delta(1,1')\\ &+i\hbar G(1,1')W(1',1)\,.
    \label{eq:selfenergy}
\end{split}
\end{equation}
In a quasi-particle picture, it follows that renormalized energies are given by the self-consistency relation
\begin{equation}
\begin{split} 
  E^{\lambda}_{\bk}&=\varepsilon_{\bk}^{\lambda}+\Sigma_{\bk}^{\textrm{H},\lambda}+\textrm{Re}\,\Sigma_{\bk}^{\textrm{GW},\textrm{ret},\lambda}(E^\lambda_{\bk})\\
            &=\varepsilon_{\bk}^{\lambda}+\Sigma_{\bk}^{\textrm{H},\lambda}+\Sigma_{\bk}^{\textrm{F},\lambda}+\textrm{Re}\,\Sigma_{\bk}^{\textrm{MW},\textrm{ret},\lambda}(E^{\lambda}_{\bk}),
    \label{eq:GW_energy}
\end{split}
\end{equation}
where the self-energies describe photoexcited carriers in the electron-hole picture. The corresponding quasi-particle dampings are 
$\Gamma^{\lambda}_{\bq}=-\textrm{Im}\,\Sigma_{\bk}^{\textrm{MW},\textrm{ret},\lambda}(E^{\lambda}_{\bk})$.
We have split the GW self-energy into an instantaneous Fock term and the so-called Montroll-Ward term according to the decomposition of the 
retarded screened Coulomb interaction matrix \cite{kremp_quantum_2005}:
\begin{equation}
\begin{split} 
  W^{\textrm{ret}}_{\alpha\beta,\bq}(t)=W^{\delta}_{\alpha\beta,\bq}\delta(t)+\theta(t)\Big[W^{>}_{\alpha\beta,\bq}(t) - W^{<}_{\alpha\beta,\bq}(t) \Big]\,.
    \label{eq:W_ret}
\end{split}
\end{equation}
Note that the Coulomb interaction $W$ contains both, background screening from the semiconductor and its dielectric environment in the ground state and screening from photoexcited carriers \cite{erben_excitation-induced_2018}. 
The Montroll-Ward self-energy is explicitly given by:
\begin{equation}
\begin{split} 
  & \Sigma_{\bk}^{\textrm{MW},\textrm{ret},\lambda}(\omega) = i\hbar\int_{-\infty}^{\infty}\frac{d\omega'}{2\pi}\\
            \frac{1}{\mathcal{A}}\sum_{\bk'\lambda'}&\frac{(1-f^{\lambda'}_{\bk'})W^{>,\lambda'\lambda\lambda'\lambda}_{\bk'\bk\bk'\bk}(\omega')+ 
            f^{\lambda'}_{\bk'}W^{<,\lambda'\lambda\lambda'\lambda}_{\bk'\bk\bk'\bk}(\omega')}{\hbar\omega-E^{\lambda'}_{\bk'}+i\Gamma^{\lambda'}_{\bk'}-\hbar\omega'}
     \,,
    \label{eq:MW}
\end{split}
\end{equation}
with the Fermi distribution functions for electrons and holes $f^{\lambda}_{\bk}$ and the crystal area $\mathcal{A}$. The band sum is limited such that electron-hole exchange is not taken into account.
The plasmon propagators in Bloch representation are connected to the Coulomb matrix by:
\begin{equation}
\begin{split} 
  W^{\gtrless,\lambda_1\lambda_2\lambda_3\lambda_4}_{\bk_1\bk_2\bk_3\bk_4}(\omega)=\sum_{\alpha,\beta} 
  (c^{\lambda_1}_{\alpha,\bk_1})^*(c^{\lambda_2}_{\beta,\bk_2})^*c^{\lambda_3}_{\beta,\bk_3}c^{\lambda_4}_{\alpha,\bk_4}
  W^{\gtrless}_{\alpha\beta,\bk_3-\bk_2}(\omega)
     \,.
    \label{eq:W_prop_Bloch}
\end{split}
\end{equation}
In quasi-equilibrium, the propagators fulfill the Kubo-Martin-Schwinger relation \cite{kremp_quantum_2005}
\begin{equation}
\begin{split} 
  W^{>}_{\alpha\beta,\bq}(\omega)= e^{\frac{\hbar\omega}{k_{\textrm{B}} T}} W^{<}_{\alpha\beta,\bq}(\omega)\,.
     \,
    \label{eq:W_prop_KMS}
\end{split}
\end{equation}
Combining this with the Kramers-Kronig relations for the inverse dielectric function (see below) and Eq.~(\ref{eq:W_ret}),
the propagators can be expressed in terms of the retarded Coulomb interaction:
\begin{equation}
\begin{split} 
  W^{>}_{\alpha\beta,\bq}(\omega)&=(1+n_{\textrm{B}}(\omega))\,2i\,\textrm{Im}\,W^{\textrm{ret}}_{\alpha\beta,\bq}(\omega), \\
  W^{<}_{\alpha\beta,\bq}(\omega)&=   n_{\textrm{B}}(\omega) \,2i\,\textrm{Im}\,W^{\textrm{ret}}_{\alpha\beta,\bq}(\omega)
     \,
    \label{eq:W_prop_ret}
\end{split}
\end{equation}
with the Bose distribution function $n_{\textrm{B}}(\omega)$. The retarded Coulomb matrix is obtained using the dielectric matrix
for photoexcited carriers:
\begin{equation}
\begin{split} 
  W^{\textrm{ret}}_{\alpha\beta,\bq}(\omega)&=\sum_{\gamma} \varepsilon^{-1,\textrm{ret},\alpha\gamma}_{\textrm{exc},\bq}(\omega)V^{\textrm{ret}}_{\gamma\beta,\bq}(\omega)\,.
    \label{eq:W_ret_eps}
\end{split}
\end{equation}
The dielectric matrix itself is given by
\begin{equation}
	\begin{split} 
		\varepsilon^{\textrm{ret},\alpha\beta}_{\textrm{exc},\bq}(\omega)=\delta_{\alpha\beta}-\sum_{\gamma}V^{\alpha\gamma}_{\bq} P^{\gamma\beta}_{\textrm{exc},\bq}(\omega)\,,
		\label{eq:eps_ab}
	\end{split}
\end{equation}
where we describe the polarization matrix for photoexcited carriers in the Lindhard (random-phase) approximation
\begin{equation} 
	\begin{split}  
		P^{\alpha\beta}_{\textrm{exc},\bq}(\omega)&=\frac{1}{\mathcal{A}}\sum_{\lambda,\lambda',\bk} c_{\alpha, \mathbf{k}}^{\lambda} c_{\beta, \mathbf{k}-\mathbf{q}}^{\lambda'}  \Big(c_{\beta, \mathbf{k}}^{\lambda} \Big)^{*} \Big(c_{\alpha, \mathbf{k}-\mathbf{q}}^{\lambda'} \Big)^{*} \\
		&\times\frac{f_{\bk-\bq}^{\lambda'}-f_{\bk}^{\lambda}}{\varepsilon_{\bk-\bq}^{\lambda'}-\varepsilon_{\bk}^{\lambda}+\hbar\omega+i\gamma}~.
		\label{eq:lindhard}
	\end{split}
\end{equation}
We use a phenomenological damping $\gamma =$ min($10$ meV, $\hbar\omega$) to ensure the correct analytic behavior in the static limit $\omega\rightarrow 0$.
In summary, the Montroll-Ward self-energy is given by
\begin{equation}
	\begin{split} 
		\Sigma_{\bk}^{\textrm{MW},\textrm{ret},\lambda}(\omega) & = i\hbar\int_{-\infty}^{\infty}\frac{d\omega'}{2\pi}
		\frac{1}{\mathcal{A}}\sum_{\bk'\lambda'}\sum_{\alpha\beta}
		(c^{\lambda}_{\alpha,\bk})^*(c^{\lambda'}_{\beta,\bk'})^*c^{\lambda}_{\beta,\bk}c^{\lambda'}_{\alpha,\bk'} \\
		&\times 2i\,\textrm{Im}\,\Big\{ W^{\textrm{ret}}_{\alpha\beta,\bk-\bk'}(\omega')\Big\}
		\frac{1-f^{\lambda'}_{\bk'}+n_{\textrm{B}}(\omega')}{\hbar\omega-E^{\lambda'}_{\bk'}+i\Gamma^{\lambda'}_{\bk'}-\hbar\omega'}
		\,.
		\label{eq:MW_final}
	\end{split}
\end{equation}
\subsection{Ground-state renormalizations due to substrate screening}\label{sec:ground_state}

We note that even in the absence of photoexcited carriers, i.e. $P^{\alpha\beta}_{\textrm{exc},\bq}(\omega)=0$ and $\varepsilon^{\textrm{ret},\alpha\beta}_{\textrm{exc},\bq}(\omega)=\delta_{\alpha\beta}$,
the Montroll-Ward self-energy yields a finite contribution due to the imaginary part of the frequency-dependent background-screened matrix element
\begin{equation}
\begin{split} 
  V^{\textrm{ret}}_{\alpha\beta,\bq}(\omega)=\sum_{\gamma} \varepsilon^{-1,\textrm{ret},\alpha\gamma}_{\textrm{b},\bq}(\omega)U_{\gamma\beta,\bq}\,.
    \label{eq:V_mat}
\end{split}
\end{equation}
These quasi-particle renormalizations due to dynamical substrate screening can be interpreted as a correction to the band structure $\varepsilon_{\bk}^{\lambda}$ of the freestanding TMD layer in the sense of a GdW calculation \cite{rohlfing_electronic_2010}.
Using the WFCE scheme \cite{rosner_wannier_2015}, the background dielectric matrix $\varepsilon^{-1,\textrm{ret},\alpha\beta}_{\textrm{b},\bq}(\omega)$ is obtained from the DFT-based matrix for the freestanding WS$_2$ monolayer by replacing its leading eigenvalue with a macroscopic dielectric function for the desired van der Waals heterostructure. The macroscopic dielectric function is calculated by solving Poisson's equation \cite{florian_dielectric_2018} including bulk-like dielectric functions for each layer. Due to the model character of the macroscopic dielectric function for the heterostructure, care has to be taken to properly treat the static limit of background screening. While graphene $\pi$-band screening is described by a frequency-dependent polarization (see below), screening from higher graphene bands, from the WS$_2$ itself, as well as from other components of the heterostructure is assumed to be static.
In general, the background dielectric matrix is composed of a static and a dynamical part:
\begin{equation}
\begin{split} 
  \varepsilon^{-1,\textrm{ret},\alpha\beta}_{\textrm{b},\bq}(\omega)=\varepsilon^{-1,\textrm{stat},\alpha\beta}_{\textrm{b},\bq}+\varepsilon^{-1,\textrm{dyn},\alpha\beta}_{\textrm{b},\bq}(\omega)\,.
    \label{eq:eps_b_stat_dyn}
\end{split}
\end{equation}
A proper dielectric function fulfilling the Kramers-Kronig relation
\begin{equation}
\begin{split} 
  \textrm{Re}\,\varepsilon^{-1,\textrm{ret},\alpha\beta}_{\textrm{b},\bq}(\omega)=\delta_{\alpha\beta}+\mathcal{P}\,\int_{-\infty}^{\infty}\frac{d\omega'}{\pi}
  \frac{\textrm{Im}\,\varepsilon^{-1,\textrm{ret},\alpha\beta}_{\textrm{b},\bq}(\omega')}{\omega'-\omega}
    \label{eq:eps_KK}
\end{split}
\end{equation}
would have $\varepsilon^{-1,\textrm{stat},\alpha\beta}_{\textrm{b},\bq}=\delta_{\alpha\beta}$. A model dielectric function, in the simplest case describing the static limit by a constant $\varepsilon_{\infty}$, leads to a violation of this behavior of the form $\varepsilon^{-1,\textrm{stat},\alpha\beta}_{\textrm{b},\bq}=\delta_{\alpha\beta}+X^{\alpha\beta}_{\bq}$. The static contribution can be extracted from the full dielectric function as
\begin{equation}
\begin{split} 
  \varepsilon^{-1,\textrm{stat},\alpha\beta}_{\textrm{b},\bq}=\textrm{Re}\,\varepsilon^{-1,\textrm{ret},\alpha\beta}_{\textrm{b},\bq}(0)
  -\mathcal{P}\,\int_{-\infty}^{\infty}\frac{d\omega'}{\pi}
  \frac{\textrm{Im}\,\varepsilon^{-1,\textrm{ret},\alpha\beta}_{\textrm{b},\bq}(\omega')}{\omega'}\,,
    \label{eq:eps_static}
\end{split}
\end{equation}
yielding a statically screened matrix element 
\begin{equation}
\begin{split} 
  V^{\textrm{stat},\alpha\beta}_{\bq}=\sum_{\gamma} \varepsilon^{-1,\textrm{stat},\alpha\gamma}_{\textrm{b},\bq} U_{\gamma\beta,\bq}\,.
    \label{eq:V_stat_mat}
\end{split}
\end{equation}
Since the static contribution has no imaginary part, it does not contribute to the Montroll-Ward renormalization (\ref{eq:MW_final}). While this is natural for a proper dielectric function, where all non-trivial contributions are dynamical (i.e. $\varepsilon^{-1,\textrm{stat},\alpha\beta}_{\textrm{b},\bq}=\delta_{\alpha\beta}$), we have to include the static contribution to band-structure renormalizations explicitly as an extra term. This term is derived by taking the static limit of the Montroll-Ward term (\ref{eq:MW_final}) for $W^{\textrm{ret}}_{\alpha\beta,\bq}(\omega)=V^{\textrm{ret}}_{\alpha\beta,\bq}(\omega)$ along the lines of Ref.~\cite{erben_excitation-induced_2018}, which yields a Screened-Exchange-Coulomb-Hole self-energy. However, as discussed above, we replace $V^{\textrm{ret}}_{\alpha\beta,\bq}(\omega=0)$ by the modified static matrix $V^{\textrm{stat},\alpha\beta}_{\bq}$ as the static limit of screened Coulomb interaction. Hence, in the Bloch representation, we have the screened matrix elements 
\begin{equation}
\begin{split} 
  V^{\textrm{stat},\lambda\lambda'\lambda\lambda'}_{\bk\bk'\bk\bk'}=\sum_{\alpha,\beta} 
  (c^{\lambda}_{\alpha,\bk})^*(c^{\lambda'}_{\beta,\bk'})^*c^{\lambda}_{\beta,\bk}c^{\lambda'}_{\alpha,\bk'}
  V^{\textrm{stat},\alpha\beta}_{\bk-\bk'}
    \label{eq:V_stat_Bloch}
\end{split}
\end{equation}
and the bare matrix elements
\begin{equation}
\begin{split} 
  U^{\lambda\lambda'\lambda\lambda'}_{\bk\bk'\bk\bk'}=\sum_{\alpha,\beta} 
  (c^{\lambda}_{\alpha,\bk})^*(c^{\lambda'}_{\beta,\bk'})^*c^{\lambda}_{\beta,\bk}c^{\lambda'}_{\alpha,\bk'}
  U^{\alpha\beta}_{\bk-\bk'}\,.
    \label{eq:U_Bloch}
\end{split}
\end{equation}
The resulting static self-energy is
\begin{equation}
\begin{split} 
    \Sigma_{\bk}^{\textrm{\textrm{MW},stat},\lambda} &= -\frac{1}{\mathcal{A}}\sum_{\bk'\lambda'} V^{\textrm{stat},\lambda\lambda'\lambda\lambda'}_{\bk\bk'\bk'\bk'}
    f^{\lambda'}_{\bk'} \\
    &+\frac{1}{2\mathcal{A}}\sum_{\bk'\lambda'}(V^{\textrm{stat},\lambda\lambda'\lambda\lambda'}_{\bk\bk'\bk'\bk'}
    -U^{\lambda\lambda'\lambda\lambda'}_{\bk\bk'\bk\bk'})\,.
    \label{eq:MW_stat}
\end{split}
\end{equation}
To avoid double-counting, we have to subtract static contributions from the ground-state ($f^{\lambda}_{\bk}=0$) freestanding WS$_2$ monolayer, which are already contained in the GW-based band structure $\varepsilon_{\bk}^{\lambda}$:
\begin{equation}
\begin{split} 
    \Sigma_{\bk}^{\textrm{freest.},\lambda} = \frac{1}{2\mathcal{A}}\sum_{\bk'\lambda'}(V^{\textrm{freest.},\lambda\lambda'\lambda\lambda'}_{\bk\bk'\bk'\bk'}
    -U^{\lambda\lambda'\lambda\lambda'}_{\bk\bk'\bk\bk'})\,.
    \label{eq:freest_stat}
\end{split}
\end{equation}
Note that the matrix elements $U$ always describe bare Coulomb interaction in the freestanding monolayer. The total static self-energy is therefore given by:
\begin{equation}
\begin{split} 
    \Sigma_{\bk}^{\textrm{stat},\lambda} &= \Sigma_{\bk}^{\textrm{\textrm{MW},stat},\lambda} - \Sigma_{\bk}^{\textrm{freest.},\lambda} \\
    &=-\frac{1}{\mathcal{A}}\sum_{\bk'\lambda'} V^{\textrm{stat},\lambda\lambda'\lambda\lambda'}_{\bk\bk'\bk'\bk'}
    f^{\lambda'}_{\bk'} \\
    &+\frac{1}{2\mathcal{A}}\sum_{\bk'\lambda'}(V^{\textrm{stat},\lambda\lambda'\lambda\lambda'}_{\bk\bk'\bk'\bk'}
    -V^{\textrm{freest.},\lambda\lambda'\lambda\lambda'}_{\bk\bk'\bk\bk'})\,.
    \label{eq:GdW_stat}
\end{split}
\end{equation}
The second term has the form of a static GdW self-energy, with a Coulomb potential given by the difference between a strongly and a weakly screened potential.

\subsection{Fock self-energy with static background screening}

In the same spirit as for the ground-state contribution of screening, the static interaction $W^{\delta}_{\alpha\beta,\bq}$ in Eq.~(\ref{eq:W_ret}) is described by the corrected Coulomb matrix $V^{\textrm{stat},\alpha\beta}_{\bq}$ given by Eq.~(\ref{eq:V_stat_mat}) to obtain the Fock self-energy for excited carriers:
\begin{equation}
\begin{split} 
    \Sigma_{\bk}^{\textrm{F},\lambda} =& -\frac{1}{\mathcal{A}}\sum_{\bk'\lambda'} V^{\textrm{stat},\lambda\lambda'\lambda\lambda'}_{\bk\bk'\bk\bk'} f^{\lambda'}_{\bk'} \\
                                      &+\frac{1}{\mathcal{A}}\sum_{\bk'\bar{\lambda}'} U^{\lambda\bar{\lambda}'\lambda\bar{\lambda}'}_{\bk\bk'\bk\bk'} f^{\bar{\lambda}'}_{\bk'} \,.
    \label{eq:Fock}
\end{split}
\end{equation}
Here, the index $\lambda'$ runs over bands within the same carrier species as $\lambda$, while $\bar{\lambda}'$ sums over carriers with opposite charge.
As discussed in \cite{erben_excitation-induced_2018}, electron-hole exchange is described by unscreened Coulomb matrix elements.

\subsection{Hartree self-energy with frequency-dependent background screening}

In the following, we derive an expression for the Hartree self-energy using the above separation of background screening into static and dynamical parts.
In the Bloch basis, the self-energy in (\ref{eq:selfenergy}) takes the form:
\begin{equation}
\begin{split} 
    &\Sigma_{\bk}^{\textrm{H},\lambda}(t,t')  \\
    &= -i\hbar\frac{1}{\mathcal{A}}\sum_{\bk'\lambda'}\int_{-\infty}^{\infty} dt_2 V^{\textrm{ret},\lambda\lambda'\lambda'\lambda}_{\bk\bk'\bk'\bk}(t,t_2) G^{<,\lambda'}_{\bk'}(t_2,t_2^+)\delta(t-t') \\
    &=\frac{1}{\mathcal{A}}\sum_{\bk'\lambda} \sum_{\alpha\beta}(c^{\lambda}_{\alpha,\bk})^*(c^{\lambda'}_{\beta,\bk'})^*c^{\lambda'}_{\beta,\bk'}c^{\lambda}_{\alpha,\bk}\\
    &\times\int_{-\infty}^{\infty} dt_2 V^{\textrm{ret}}_{\alpha\beta,\boldsymbol{0}}(t,t_2) f^{\lambda'}_{\bk'}(t_2)\delta(t-t')
    \,,
    \label{eq:Hartree}
\end{split}
\end{equation}
where we have replaced the equal-time propagator $-i\hbar G^<(t,t^+)$ by single-particle occupations $f(t)$. Assuming a quasi-equilibrium state, we drop the time dependence of the latter, which leads to a time integral
\begin{equation}
\begin{split} 
    \int_{-\infty}^{\infty} dt_2 V^{\textrm{ret}}_{\alpha\beta,\boldsymbol{0}}(t,t_2)=
    \sum_{\gamma}\int_{-\infty}^{\infty} dt_2 \varepsilon^{-1,\textrm{ret},\alpha\gamma}_{\textrm{b},\boldsymbol{0}}(t,t_2)U_{\gamma\beta,\boldsymbol{0}}
    \,.
    \label{eq:V_time_int}
\end{split}
\end{equation}
As in Eq.~(\ref{eq:eps_b_stat_dyn}), we spilt the dielectric function into an instantaneous (static) and a retarded (dynamical) part:
\begin{equation}
\begin{split} 
    &\varepsilon^{-1,\textrm{ret},\alpha\gamma}_{\textrm{b},\boldsymbol{0}}(t,t_2)= \\
    &\delta(t-t_2)\varepsilon^{-1,\textrm{stat},\alpha\gamma}_{\textrm{b},\boldsymbol{0}}+
    \theta(t-t_2)\varepsilon^{-1,\textrm{dyn},\alpha\gamma}_{\textrm{b},\boldsymbol{0}}(t,t_2)
    \,.
    \label{eq:eps_b_inst_ret}
\end{split}
\end{equation}
Expressing the retarded part in quasi-equilibrium as 
\begin{equation}
\begin{split} 
    &\varepsilon^{-1,\textrm{dyn},\alpha\gamma}_{\textrm{b},\boldsymbol{0}}(\tau) \\
    =&\lim_{\eta\rightarrow 0^+}\int_{-\infty}^{\infty}\frac{d\omega}{2\pi}e^{-i(\omega-i\eta)\tau}
    (\varepsilon^{-1,\textrm{ret},\alpha\gamma}_{\textrm{b},\boldsymbol{0}}(\omega)
    -\varepsilon^{-1,\textrm{stat},\alpha\gamma}_{\textrm{b},\boldsymbol{0}})
    \,
    \label{eq:eps_b_ret_eq}
\end{split}
\end{equation}
and using the modified Kramers-Kronig relation (\ref{eq:eps_static}), we arrive at
\begin{equation}
\begin{split} 
    \int_{-\infty}^{\infty} dt_2 \varepsilon^{-1,\textrm{ret},\alpha\gamma}_{\textrm{b},\boldsymbol{0}}(t,t_2)=\textrm{Re}\,\varepsilon^{-1,\textrm{ret},\alpha\gamma}_{\textrm{b},\boldsymbol{0}}(\omega=0)
    \,.
    \label{eq:eps_time_int}
\end{split}
\end{equation}
Therefore, the Hartree interaction is described by the zero-frequency limit of the background-screened Coulomb potential:
\begin{equation}
\begin{split} 
    \Sigma_{\bk}^{\textrm{H},\lambda} =& \frac{1}{\mathcal{A}}\sum_{\bk'\lambda'} V^{\textrm{ret},\lambda\lambda'\lambda'\lambda}_{\bk\bk'\bk'\bk}(\omega=0) f^{\lambda'}_{\bk'}  \\
                                      -& \frac{1}{\mathcal{A}}\sum_{\bk'\bar{\lambda}'} V^{\textrm{ret},\lambda\bar{\lambda}'\bar{\lambda}'\lambda}_{\bk\bk'\bk'\bk}(\omega=0) f^{\bar{\lambda}'}_{\bk'} \,.
    \label{eq:Hartree}
\end{split}
\end{equation}

\subsection{Hartree interaction in Wannier representation}

Due to the Coulomb singularity at long wavelength ($\bq=0$), the Hartree interaction requires a separate treatment in the Wannier representation \cite{schobert_ab_2024}. 
The corresponding matrix element is given by 
\begin{equation}
\begin{split} 
    &V^{\textrm{ret},\lambda\lambda'\lambda'\lambda}_{\bk\bk'\bk'\bk}(\omega=0) \\ 
    =&\sum_{\alpha\beta} (c^{\lambda}_{\alpha,\bk})^*(c^{\lambda'}_{\beta,\bk'})^*c^{\lambda'}_{\beta,\bk'}c^{\lambda}_{\alpha,\bk}
    V^{\textrm{ret}}_{\alpha\beta,\boldsymbol{0}}(\omega=0) \\
    =&\sum_{\alpha\beta} |c^{\lambda}_{\alpha,\bk}|^2|c^{\lambda'}_{\beta,\bk'}|^2
    \sum_{\gamma\delta} T_{\alpha\gamma} V^{\textrm{D}}_{\gamma\delta,\boldsymbol{0}}(\omega=0) T^{\dagger}_{\delta\beta}
    \,,
    \label{eq:Hartree_ME}
\end{split}
\end{equation}
where the columns of $T_{\alpha\gamma}$ are the eigenvectors of the Coulomb matrix (assumed to be momentum- and frequency-independent) and
$V^{\textrm{D}}_{\gamma\delta,\boldsymbol{0}}(\omega=0)$ is a diagonal matrix composed of the corresponding eigenvalues $V_{\gamma,\boldsymbol{0}}(\omega=0)$. As discussed in \cite{rosner_wannier_2015},
the leading eigenvalue describes the macroscopic, or long-range properties of the Coulomb interaction, while the other eigenvalues are responsible for microscopic effects within the crystal unit cell. We can therefore decompose the Hartree self-energy as
\begin{equation}
\begin{split} 
    &\Sigma_{\bk}^{\textrm{H},\lambda}\\ 
    =&\frac{1}{\mathcal{A}}\sum_{\bk'\lambda'} \Big(V_{1,\boldsymbol{0}}(0) \sum_{\alpha\beta} |c^{\lambda}_{\alpha,\bk}|^2|c^{\lambda'}_{\beta,\bk'}|^2
    T_{\alpha,1} T^{\dagger}_{1,\beta}\\
    &+V^{\textrm{micro},\lambda\lambda'\lambda'\lambda}_{\bk\bk'\bk'\bk}(0) \Big) f^{\lambda'}_{\bk'} s^{\lambda\lambda'}
    \,
    \label{eq:Hartree_decompose}
\end{split}
\end{equation}
with $s^{\lambda\lambda'}$ taking into account the sign of the Coulomb interaction term, being $+1$ for like charges and $-1$ for unlike charges.
The macroscopic eigenvector has the same contribution for each of the $n_{\textrm{orb}}$ orbitals, $\boldsymbol{T_{1}}=\frac{1}{\sqrt{n_{\textrm{orb}}}}(1,1,...,1)$, while the macroscopic 
eigenvalue is proportional to the number of orbitals: $V_{1,\boldsymbol{\bq}}(0)=n_{\textrm{orb}} \frac{e^2}{2\varepsilon_0 q}\tilde{V}_q $ with a factor $\tilde{V}_q$ describing the non-trivial momentum dependence. Normalization of the Bloch states
($\sum_{\alpha} |c^{\lambda}_{\alpha,\bk}|^2=1 $) then leads to
\begin{equation}
\begin{split} 
    \Sigma_{\bk}^{\textrm{H},\lambda}
    =\lim_{q\rightarrow 0}\sum_{\lambda'}s^{\lambda\lambda'}n^{\lambda'} \frac{e^2}{2\varepsilon_0 q}\tilde{V}_0
    +\Sigma_{\bk}^{\textrm{H, micro},\lambda}
    \,
    \label{eq:Hartree_decompose}
\end{split}
\end{equation}
with the carrier density $n^{\lambda}=\frac{1}{\mathcal{A}}\sum_{\bk}f^{\lambda}_{\bk}$ in band $\lambda$. For a globally charge-neutral system, the macroscopic term drops out and only microscopic contributions to Hartree interaction remain. In a heterostructure exhibiting charge transfer, as in the present case of WS$_2$ on graphene, the individual layers become globally charged. Here, we do not attempt to quantify the corresponding charging shifts from theory but subtract them from the experimental data for direct theory-experiment comparison. Hence we calculate Hartree-type matrix elements by setting the macroscopic eigenvalue to zero before transforming the matrix element back to the Wannier representation. Note that since only the macroscopic eigenvalue is modified by environmental screening according to the WFCE scheme, Hartree interaction is not influenced by the graphene substrate but only by background screening within the WS$_2$ unit cell.

\subsection{Frequency-dependent screening from graphene substrate}\label{sec:graphene}
\begin{figure}[h!t]
\centering
\includegraphics[width=0.7\columnwidth]{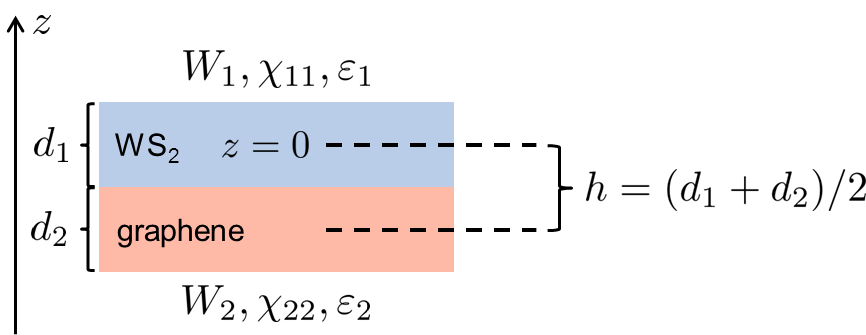}
\caption{Schematic of the WS$_2$/graphene heterostructure. $W_i$, $\chi_{ii}$ and $\varepsilon_i$ denote the screened interaction, polarization and dielectric function in the respective layers. The $z$ coordinate is centered at the WS$_2$-layer.} 
\label{fig:WS$_2$_gr_HS}
\end{figure}
As discussed above, we describe environmental screening of Coulomb interaction in a two-dimensional layer, in particular due to a single layer of graphene below, in terms of a macroscopic dielectric function obtained from Poisson's equation for a given heterostructure \cite{florian_dielectric_2018}. This approach assumes that a single layer of graphene can be modeled as a dielectric slab with a certain effective width $d_2$ and an effective bulk-like dielectric function $\varepsilon_2$, see Fig.~\ref{fig:WS$_2$_gr_HS}. To obtain accurate results, it is essential that the efficient metallic screening of graphene in the long-wavelength limit is captured by the the analytic form of $\varepsilon_2$. The goal of this section is to derive such an effective dielectric function from the microscopic two-dimensional polarization $\chi_{22} $ of graphene.
\\
\\We start along the lines of \cite{lin_plasmons_1997} by setting up a Dyson-type equation for the screened interactions $W_1$ and $W_2$ in the WS$_2$ and graphene layers, which are at a distance $h$:
\begin{equation}
\begin{split} 
    \varepsilon_b 
    \begin{pmatrix}
      W_1 \\
      W_2
    \end{pmatrix}
    =
     \begin{pmatrix}
      V^{\textrm{ex}}_1 \\
      V^{\textrm{ex}}_2
    \end{pmatrix}
    +
    \begin{pmatrix}
      V_{11} & V_{12}   \\
      V_{21} & V_{22}
    \end{pmatrix}
    \begin{pmatrix}
      \chi_{11} & 0   \\
      0 & \chi_{22}   
    \end{pmatrix}
    \begin{pmatrix}
      W_1 \\
      W_2
    \end{pmatrix}
    \,.
    \label{eq:dyson_W}
\end{split}
\end{equation}
This Dyson-type approach is frequently used to describe screening in graphite, modeling the individual layers as truly two-dimensional systems with corresponding polarizations. In this sense, the approach is complementary to Poisson's equation, where each layer is effectively modeled as a three-dimensional object with bulk-like dielectric function.
In Eq.~(\ref{eq:dyson_W}), $\varepsilon_b$ is a background dielectric constant, $V^{\textrm{ex}}_i$ are the potentials generated by ``external'' charges in the layers, $V_{ij}$ is the bare Coulomb interaction between layers $i$ and $j$ and $\chi_{ii}$ is the irreducible polarization of layer $i$. Placing a test charge in the WS$_2$ layer, it is $V^{\textrm{ex}}_1=V_q$ and $V^{\textrm{ex}}_2=V_q e^{-hq}$. Similary, the bare Coulomb interaction is $V_{ii}=V_q$ and $V_{12}=V_{21}=V_q e^{-hq}$. Neglecting the polarization of WS$_2$, we obtain
\begin{equation}
\begin{split} 
   W_1=\frac{V_q}{\varepsilon_b}\Big(1+\frac{V_q e^{-2hq}\chi_{22}}{\varepsilon_b - V_q\chi_{22}} \Big)=\frac{V_q}{\varepsilon_b}\varepsilon^{-1,\textrm{Dyson}}_q
    \,
    \label{eq:W_1}
\end{split}
\end{equation}
and
\begin{equation}
\begin{split} 
   W_2=\frac{V_q e^{-hq}}{\varepsilon_b - V_q\chi_{22}}
    \,.
    \label{eq:W_2}
\end{split}
\end{equation}
Taking into account the linear behavior of the graphene $\pi$-band polarization at long wavelengths \cite{hwang_dielectric_2007}, $\chi_{22}\approx-\alpha q$, 
we obtain in the $q=0$-limit:
\begin{equation}
\begin{split} 
   \lim_{q\rightarrow 0} \varepsilon^{\textrm{Dyson}}_q=\frac{\varepsilon_b+\tilde{\alpha}}{\varepsilon_b},\quad \tilde{\alpha}=\frac{e^2}{8\hbar v_{\textrm{F}}\varepsilon_0}
    \label{eq:eps_dyson_limit}
\end{split}
\end{equation}
with $\tilde{\alpha}\approx 3-4 $ depending on the Fermi velocity $v_{\textrm{F}}$ \cite{lin_plasmons_1997,hwang_dielectric_2007,trevisanutto_ab_2008}. Hence, unlike a two-dimensional semiconductor, a single graphene layer yields an effective dielectric constant larger than unity in the long-wavelength limit. A detailed discussion of this property is given in \cite{hwang_dielectric_2007}.
\\
\\To draw the connection to a bulk-like effective dielectric function for graphene, we solve Poisson's equation for the heterostructure shown in Fig.~\ref{fig:WS$_2$_gr_HS} for $\varepsilon_1=1$ (corresponding to zero WS$_2$ polarization), which yields the following dielectric function for charges at $z=0$:
\begin{equation}
\begin{split} 
   \varepsilon^{\textrm{Poisson}}_q=\frac{1-\tilde{\varepsilon}^2 e^{-2qd_2}}{1-\tilde{\varepsilon}^2 e^{-2qd_2} + \tilde{\varepsilon}e^{-qd_1}(e^{-2qd_2}-1)}
    \label{eq:Poisson_eps}
\end{split}
\end{equation}
with $\tilde{\varepsilon}=\frac{\varepsilon_2 - 1}{\varepsilon_2 + 1} $. More generally, Poisson's equation provides the screened potential at any position $z$ as superposition of potentials generated by the test charge at $z=0$ and surface charges at the graphene boundaries $j=1,2$ at $z_1=-d_1/2$, $z_2= -d_1/2-d_2$:
\begin{equation}
\begin{split} 
   W_{q}(z)=Ae^{-q|z|}+\sum_j B_j e^{-q|z-z_j|}\,,
    \label{eq:Poisson_pot}
\end{split}
\end{equation}
where $A=V_q/\varepsilon_b$. A straightforward calculation yields for the coefficients $B_j$:
\begin{equation}
\begin{split} 
   B_1&=\frac{A}{1-\tilde{\varepsilon}^2 e^{-2qd_2}}\tilde{\varepsilon}e^{-qd_1/2}(\tilde{\varepsilon}e^{-2qd_2}-1)\,,\\
   B_2&=\frac{A}{1-\tilde{\varepsilon}^2 e^{-2qd_2}}\tilde{\varepsilon}e^{-q(d_1/2+d_2)}(1-\tilde{\varepsilon})\,.
    \label{eq:Poisson_coeff}
\end{split}
\end{equation}
Hence we obtain for the total surface charge potential of graphene:
\begin{equation}
\begin{split} 
   B&=B_1+B_2 \\ &=\frac{A}{1-\tilde{\varepsilon}^2 e^{-2qd_2}}\tilde{\varepsilon}e^{-qd_1/2}
   \Big(\tilde{\varepsilon}e^{-2qd_2}-1 + e^{-q d_2} - \tilde{\varepsilon} e^{-q d_2}   \Big) \\ 
   &=A\tilde{\varepsilon}e^{-qd_1/2} \frac{\big( 1+ \tilde{\varepsilon}e^{-qd_2}\big)\big( e^{-q d_2} - 1 \big)}{1-\tilde{\varepsilon}^2 e^{-2qd_2}} \\
   &=A\tilde{\varepsilon}e^{-qd_1/2} \frac{e^{-q d_2} - 1 }{1-\tilde{\varepsilon} e^{-qd_2}}\,.
    \label{eq:Poisson_coeff_tot}
\end{split}
\end{equation}
The limit of an infinitely thin graphene slab is calculated by letting $d_2\rightarrow 0$ and $d_1/2\rightarrow h$:
\begin{equation}
\begin{split} 
   B\rightarrow A\tilde{\varepsilon}e^{-qh} \frac{q d_2}{\tilde{\varepsilon}(1-qd_2) - 1}\,.
    \label{eq:Poisson_coeff_tot_limit}
\end{split}
\end{equation}
Similar to Eq.~(\ref{eq:Poisson_pot}), we can interpret the screened potentials $W_1$ and $W_2$ from the Dyson-type equation (Eqs.~(\ref{eq:W_1}) and (\ref{eq:W_2})) as superpositions of a test charge at $z=0$ and a graphene surface charge at $z=-h$:
\begin{equation}
\begin{split} 
   W^{\textrm{D}}_{q}(z)=A_{\textrm{D}}e^{-q|z|}+B_{\textrm{D}} e^{-q|z+h|}\,.
    \label{eq:Dyson_pot}
\end{split}
\end{equation}
By setting $W_1=W^{\textrm{D}}_{q}(z=0)$ and $W_2=W^{\textrm{D}}_{q}(z=-h)$, we find 
\begin{equation}
\begin{split} 
   A_{\textrm{D}}&=\frac{V_q}{\varepsilon_b}, \\ B_{\textrm{D}}&=\Big(\frac{V_q}{\varepsilon_b} \Big)^2 e^{-qh}\frac{\chi_{22}}{1-V_q/\varepsilon_b \chi_{22}}
   =\Big(\frac{V_q}{\varepsilon_b} \Big)^2 e^{-qh}\tilde{\chi}_{22}\,.
    \label{eq:Dyson_coeff_tot}
\end{split}
\end{equation}
While $\chi_{22}$ is the \textit{irreducible} polarization describing the density response to the total electric field within the graphene layer, the quantity
\begin{equation}
\begin{split} 
\tilde{\chi}_{22}&=\chi_{22} (1-V_q/\varepsilon_b \chi_{22})^{-1} \\
&=\chi_{22}+\chi_{22}V_q/\varepsilon_b \chi_{22}+\chi_{22}V_q/\varepsilon_b \chi_{22}V_q/\varepsilon_b \chi_{22}+...
    \label{eq:reducible_pol}
\end{split}
\end{equation}
is the \textit{reducible} polarization responding to the external field alone.
Identifying the surface charge coefficient $B$ (\ref{eq:Poisson_coeff_tot_limit}) from Poisson's equation with the coefficient $B_{\textrm{D}}$ (\ref{eq:Dyson_coeff_tot}) from Dyson's equation for a vacuum background ($\varepsilon_b=1$) allows to derive the effective dielectric function for graphene in terms of the graphene polarization $\chi_{22}$. We find
\begin{equation}
\begin{split} 
  \tilde{\varepsilon}&=\frac{\varepsilon_2 - 1}{\varepsilon_2 + 1}=\frac{V_q \tilde{\chi}_{22}}{V_q \tilde{\chi}_{22}-q d_2(1+V_q \tilde{\chi}_{22})}\,,
    \label{eq:Poisson_Dyson_compare1}
\end{split}
\end{equation}
from which it follows that
\begin{equation}
\begin{split} 
  \varepsilon_2&=1-\frac{2}{q d_2}\frac{V_q \tilde{\chi}_{22}}{1+V_q \tilde{\chi}_{22}} \\ 
  &=1-\frac{2}{q d_2} V_q \chi_{22}\,.
    \label{eq:Poisson_Dyson_compare2}
\end{split}
\end{equation}
Note that by choosing a vacuum environment for the heterostructure and neglecting the polarization of the TMD itself, we obtained an effective bulk-like dielectric function describing the pristine screening effects of a single graphene layer.
\\
\\Finally, we introduce the microscopic irreducible polarization of graphene $\pi$-bands according to Ref.~\cite{lin_plasmons_1997}:
\begin{equation}
\begin{split} 
  &\chi^{\textrm{gr}}_{\bq}(\omega)=2\frac{1}{\mathcal{A}}\sum_{b,b'=c,v}\sum_{\bk}\frac{1}{4} \Big(1+q^2/36\Big)^{-6} \\
  \times&\Bigg|1\pm \frac{H_{12}(\bk+\bq)H^*_{12}(\bk)}{|H_{12}(\bk+\bq)H^*_{12}(\bk)|}   \Bigg|\frac{f^{b'}_{\bk+\bq}-f^{b}_{\bk}}
  {\varepsilon^{b'}_{\bk+\bq}-\varepsilon^{b}_{\bk} + \hbar\omega +i\gamma}\,,
    \label{eq:chi_gr}
\end{split}
\end{equation}
where we choose $\bq=(q,0)$ for simplicity. $+$ and $-$, respectively, correspond to intraband and interband excitations. 
$f^{b}_{\bk}$ denote Fermi distribution functions at a given temperature and chemical potential.
The dispersion is given by
\begin{equation}
\begin{split} 
  \varepsilon^{c,v}_{\bk}=\pm \gamma_0\Big[1+4\textrm{cos}(\sqrt{3} a k_y/2)\textrm{cos}(a k_x/2)+4\textrm{cos}^2(a k_x/2)\Big]^{1/2}
    \label{eq:disp_gr}
\end{split}
\end{equation}
and $\gamma_0 = 2 / \sqrt{3} \hbar v_{\textrm{F}} / a$
with the DFT-based Fermi velocity $v_{\textrm{F}}=950$ nm ps$^{-1}$ \cite{trevisanutto_ab_2008} and the lattice constant $a=0.246$ nm \cite{lin_plasmons_1997} ($\sqrt{3}$ times the C-C bond length). The matrix elements $H_{12}$ are given by
\begin{equation}
\begin{split} 
  H_{12}(\bk)=-\gamma_0\Big[e^{-i k_y a/\sqrt{3}}+2 e^{ik_y a/\sqrt{3}/2}\textrm{cos}(a k_x/2) \Big]^{1/2}\,.
    \label{eq:H_gr}
\end{split}
\end{equation}
We use a damping $\gamma =$ min($10$ meV, $\hbar\omega$) to ensure the correct analytic behavior in the static limit $\omega\rightarrow 0$.
Note that $\chi^{\textrm{gr}}_{\bq}(\omega)$ does not contain screening contributions from higher graphene bands, which are sometimes included as an additional (high-frequency) constant \cite{shung_dielectric_1986,lin_plasmons_1997}. Here, we describe the additional screening contribution by a model polarization $\chi^{\textrm{gr},\infty}_{\bq}=(1-\varepsilon^{\textrm{gr},\infty}_{\bq})/V_q$, with a dielectric function $\varepsilon^{\textrm{gr},\infty}_{\bq}$ that has been adjusted to first-principle cRPA calculations \cite{wehling_strength_2011}:
\begin{equation}
\begin{split} 
  &\varepsilon^{\textrm{gr},\infty}_{\bq}=\varepsilon_{\infty}\frac{\varepsilon_{\infty}+1-(\varepsilon_{\infty}-1)e^{-qd_{\infty}}}{\varepsilon_{\infty}+1+(\varepsilon_{\infty}-1)e^{-qd_{\infty}}} \,, \\ &\varepsilon_{\infty}=2.4\,,d_{\infty}=0.28\,\textrm{nm}\,.
    \label{eq:eps_gr_inf}
\end{split}
\end{equation}
In the following, we use the function $\varepsilon_2$ from Eq.~(\ref{eq:Poisson_Dyson_compare2}) with a standard 2-d Coulomb potential $V_q=V^{2d}_q=\frac{e^2}{2\varepsilon_0 q}$ as dielectric function of graphene in Poisson's equation. To this end, we explicitly set $\chi_{22}=\chi^{\textrm{gr}}_{\bq}(\omega)+\chi^{\textrm{gr},\infty}_{\bq}$ as irreducible polarization:
\begin{equation}
  \varepsilon_2=\varepsilon^{\textrm{gr}}_{\bq}(\omega)=1-\frac{2}{q d_{\textrm{gr}}}\frac{V^{2d}_q (\chi^{\textrm{gr}}_{\bq}(\omega)+\chi^{\textrm{gr},\infty}_{\bq})}{1-V^{2d}_q (\chi^{\textrm{gr}}_{\bq}(\omega)+\chi^{\textrm{gr},\infty}_{\bq})}\,.
    \label{eq:eps_gr_final}
\end{equation}
As graphene layer thickness $d_2$, we use the inter-layer distance in graphite $d_{\textrm{gr}}=0.335$ nm \cite{rosner_wannier_2015}.

\subsection{Results for the pristine heterostructure}

\begin{figure}[h!t]
\centering
\includegraphics[width=0.4\columnwidth]{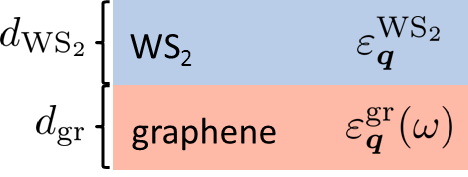}
\caption{Schematic of the WS$_2$/graphene heterostructure.} 
\label{fig:HS_pristine}
\end{figure}
\begin{figure}[h!t]
\centering
\includegraphics[width=\columnwidth]{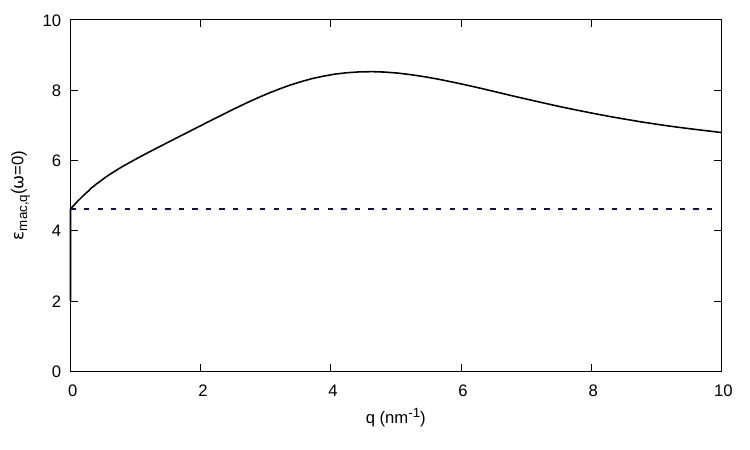}
\caption{Static limit of the macroscopic dielectric function for carriers in a WS$_2$ monolayer embedded in the heterostructure from Fig.~\ref{fig:HS_pristine} (solid line). For comparison, the analytic long-wavelength limit $1+\tilde{\alpha}$ with $\tilde{\alpha}$ from Eq.~(\ref{eq:eps_dyson_limit}) is shown as a dashed line.} 
\label{fig:eps_mac}
\end{figure}
\begin{figure}[h!t]
\centering
\includegraphics[width=\columnwidth]{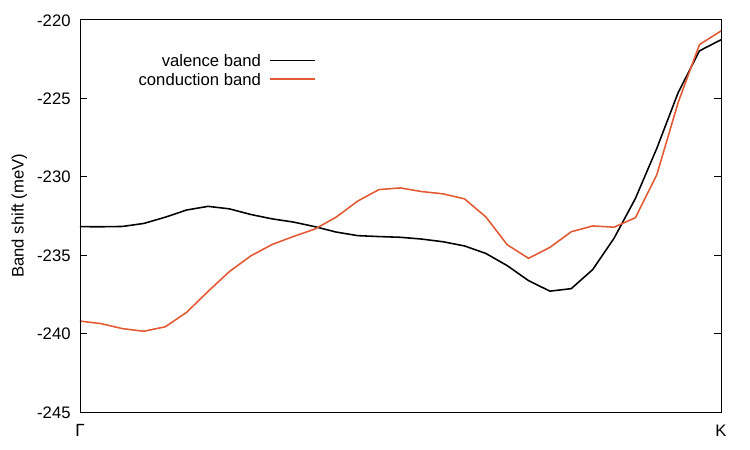}
\caption{Quasi-particle band shift of the highest valence and lowest conduction bands of monolayer WS$_2$ on an undoped graphene substrate between $\Gamma$ and K points in the absence of photoexcited carriers. The shifts are given in the electron-hole picture, i.e. both are leading to a band-gap shrinkage compared to freestanding monolayer WS$_2$.} 
\label{fig:pristine_shift_GdW}
\end{figure}

To benchmark our approach to quasi-particle renormalizations induced by screening from a graphene substrate, we evaluate the dynamical Montroll-Ward self-energy, Eq.~(\ref{eq:MW_final}), for a fixed quasi-particle broadening of $10$ meV in the absence of photoexcited carriers, i.e. $\varepsilon^{\textrm{ret},\alpha\beta}_{\textrm{exc},\bq}(\omega)=\delta_{\alpha\beta}$, $f^{\lambda}_{\bk}=0$ and $E^{\lambda}_{\bk}= \varepsilon_{\bk}^{\lambda}$. We add the corresponding static GdW contribution given by Eq.~(\ref{eq:GdW_stat}). The result is a correction to the band structure of freestanding monolayer WS$_2$:
\begin{equation}
\begin{split} 
  \Sigma_{\bk}^{\textrm{GdW},\textrm{gr},\lambda}= \Sigma_{\bk}^{\textrm{MW},\textrm{ret},\lambda}\Big|_{f^{\lambda}_{\bk}=0}(\omega=\frac{\varepsilon_{\bk}^{\lambda}}{\hbar})+\Sigma_{\bk}^{\textrm{stat},\lambda}\Big|_{f^{\lambda}_{\bk}=0}\,.
    \label{eq:Gdw_gr_pristine}
\end{split}
\end{equation}
As discussed above, screening from the dielectric environment is taken into account via the WFCE approach, replacing the leading eigenvalue of the microscopic background dielectric matrix of monolayer WS$_2$ by a macroscopic dielectric function for the given heterostructure. 
The macroscopic dielectric function is given by the Poisson solution for the heterostructure shown in Fig.~\ref{fig:HS_pristine} with $\varepsilon^{\textrm{WS}_2}_{\bq}$ and $d_{\textrm{WS}_2}$ as parametrized in Ref.~\cite{steinhoff_exciton_2017} and $\varepsilon^{\textrm{gr}}_{\bq}(\omega)$ from Eq.~(\ref{eq:eps_gr_final}). We assume zero graphene doping.
In Fig.~\ref{fig:eps_mac}, the static limit of the macroscopic dielectric function is shown. Remarkably, the long-wavelength limit is larger than $1$ due to efficient quasi-metallic screening from graphene.
\\
\\The momentum-dependent band shift resulting from Eq.~(\ref{eq:Gdw_gr_pristine}) is shown in Fig.~\ref{fig:pristine_shift_GdW}.
The renormalizations exhibit a weak band dependence, but a pronounced momentum dependence between $\Gamma$ and $K$. In the electron-hole picture, both renormalizations are negative, which means that they yield a band-gap shrinkage of about $440-480$ meV in total relative to the freestanding WS$_2$ monolayer.
%

\subsection{Photoexcited heterostructure inlcuding substrate}

\begin{figure}[h!t]
\centering
\includegraphics[width=0.4\columnwidth]{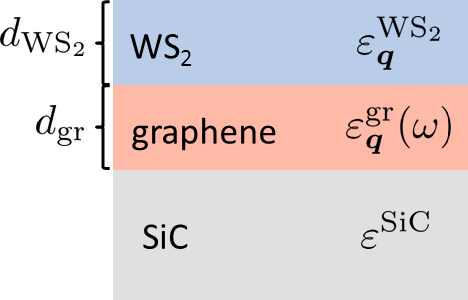}
\caption{Schematic of the WS$_2$/graphene/SiC heterostructure.} 
\label{fig:HS_full}
\end{figure}
To describe screening in the full heterostructure including a substrate, we derive a macroscopic dielectric function of the WS$_2$/Gr/SiC-heterostructure, see Fig.~\ref{fig:HS_full}. We model the pump-probe experiment assuming a sufficiently long delay after the pump pulse so that relaxation of carriers has taken place. Hence electrons and holes in the heterostructure assume a quasi-equilibrium distribution with an effective temperature that may be significantly higher than the lattice temperature. We assume that recombination of electron-hole pairs takes place on longer time scales. Prior to excitation, no carriers are excited in WS$_2$, while graphene is hole-doped with a Fermi energy $E_{\textrm{F}}=-300$ meV, corresponding to a hole density $n^{\textrm{gr,0}}_h=(E_{\textrm{F}} / \hbar / v_\textrm{F}) ^ 2 / \pi=7.3\times 10^{12}$cm$^{-2}$. Graphene carriers are assumed to be at room temperature ($T=300$ K). After optical excitation of WS$_2$ creating an electron-hole pair density $n^{\textrm{WS}_2}_e$, most of the holes are transferred to graphene as discussed in the main text. We assume that $10$ percent of the holes remain in the WS$_2$ layer, i.e. $n^{\textrm{WS}_2}_h=0.1 \times n^{\textrm{WS}_2}_e$. Correspondingly, assuming a $50$\% coverage of the graphene flakes by WS$_2$, the hole density in graphene after carrier relaxation amounts to $n^{\textrm{gr}}_h=n^{\textrm{gr,0}}_h+0.5\times0.9\times n^{\textrm{WS}_2}_e $. We assume that carriers in WS$_2$ and graphene have a common effective temperature.
\\
\\First of all, we correct the freestanding WS$_2$ band structure before the optical excitation by adding static GdW corrections due to screening from the heterostructure shown in Fig.~\ref{fig:HS_full}. To this end, we replace $\varepsilon_{\bk}^{\lambda}$ by
$\varepsilon_{\bk}^{\lambda}+\Sigma_{\bk}^{\textrm{stat},\lambda}\big|_{f^{\lambda}_{\bk}=0}$ using Eq.~(\ref{eq:GdW_stat}). Here, $\varepsilon^{\textrm{WS}_2}_{\bq}$ and $d_{\textrm{WS}_2}$ are parametrized as in Ref.~\cite{steinhoff_exciton_2017}, $\varepsilon^{\textrm{gr}}_{\bq}(\omega)$ is taken from Eq.~(\ref{eq:eps_gr_final}) and $\varepsilon^{\textrm{SiC}}=6.5 $ \cite{patrick_static_1970} is the dielectric constant of the semi-infinite substrate. Carriers in graphene are distributed as before the pump pulse.
\\
\\The pump-induced shift of WS$_2$ quasi-particle energies consists of two parts: (i) Hartree-Fock and Montroll-Ward renormalizations due to WS$_2$ carriers and (ii) screening-induced renormalizations due to graphene carriers. Here, graphene carriers are treated as ``background'' contribution to screening via the substrate dielectric function, while only carriers in WS$_2$ contribute to excited-carrier screening.
\\Renormalizations due to WS$_2$ carriers are computed by evaluating Eq.~(\ref{eq:GW_energy}) with the self-energies~(\ref{eq:MW_final}), (\ref{eq:Fock}) and (\ref{eq:Hartree}) self-consistently. Background screening is given by the doped graphene without excited carriers and excited-carrier screening due to WS$_2$ carriers is described by a dielectric matrix in RPA as given by Eqs.~(\ref{eq:eps_ab}) and (\ref{eq:lindhard}):
\begin{equation}
\begin{split} 
\Sigma_{\bk}^{\textrm{WS}_2,\lambda}&= \Sigma_{\bk,\textrm{gr doped}}^{\textrm{H},\lambda}+\Sigma_{\bk,\textrm{gr doped}}^{\textrm{F},\lambda}+\Sigma_{\bk,\textrm{gr doped}}^{\textrm{MW},\textrm{ret},\lambda}(\omega=\frac{E_{\bk}^{\lambda}}{\hbar})\,.
    \label{eq:WS2_shift}
\end{split}
\end{equation}
The Montroll-Ward term $\Sigma_{\bk,\textrm{gr doped}}^{\textrm{MW},\textrm{ret},\lambda}$ would contain dynamical background screening even in the absence of excited carriers in WS$_2$ (for $\varepsilon^{\textrm{ret},\alpha\beta}_{\textrm{exc},\bq}(\omega)=\delta_{\alpha\beta}$, see Eq.~(\ref{eq:eps_ab})) and therefore already induce some renormalizations. Hence we subtract a corresponding term without excited carriers to get a purely pump-induced shift due to WS$_2$ carriers:
\begin{equation}
\begin{split} 
\Sigma_{\bk}^{\textrm{WS}_2,\Delta,\lambda}&=\Sigma_{\bk}^{\textrm{WS}_2,\lambda}-\Sigma_{\bk}^{\textrm{WS}_2,\lambda}\Big|_{n^{\textrm{WS}_2}=0}\,.
    \label{eq:WS2_shift_diff}
\end{split}
\end{equation}
The pump-induced shift of quasi-particle energies due to graphene carriers is simulated by computing the sum of dynamical (\ref{eq:MW_final}) and static (\ref{eq:GdW_stat}) Montroll-Ward self-energies for the graphene carrier distributions before and after the pulse. For the WS$_2$ carrier populations $f^{\lambda}_{\bk}$ and quasi-particle energies $E^{\lambda}_{\bk}$, we use the values obtained from the self-consistent calculation of excited-carrier self-energies. For the excited-carrier screening, we again use $\varepsilon^{\textrm{ret},\alpha\beta}_{\textrm{exc},\bq}(\omega)=\delta_{\alpha\beta}$ to capture just the background effects of graphene. In this sense, different carrier distributions in graphene before and after the pulse induce different dielectric screening experienced by the WS$_2$ layer, leading to a differential quasi-particle renormalization.
Hence, we calculate the graphene-induced GdW shift as
\begin{equation}
\begin{split} 
\Sigma_{\bk}^{\textrm{GdW},\textrm{gr},\lambda}&= \Sigma_{\bk,\textrm{gr exc.}}^{\textrm{MW},\textrm{ret},\lambda}\Big|_{P_{\textrm{exc}}=0}(\omega=\frac{E_{\bk}^{\lambda}}{\hbar})\\ &-\Sigma_{\bk,\textrm{gr doped}}^{\textrm{MW},\textrm{ret},\lambda}\Big|_{P_{\textrm{exc}}=0}(\omega=\frac{E_{\bk}^{\lambda}}{\hbar})
\\ &+\Sigma_{\bk,\textrm{gr exc.}}^{\textrm{stat},\lambda}-\Sigma_{\bk,\textrm{gr doped}}^{\textrm{stat},\lambda}\,.
    \label{eq:Gdw_gr}
\end{split}
\end{equation}

    \newpage
    \clearpage

\subsection{Band-structure renormalizations in the photoexcited hetero\-structure for $T=1000$ K}

    \begin{figure}[h!t]
    \centering
		\includegraphics[width=.9\columnwidth]{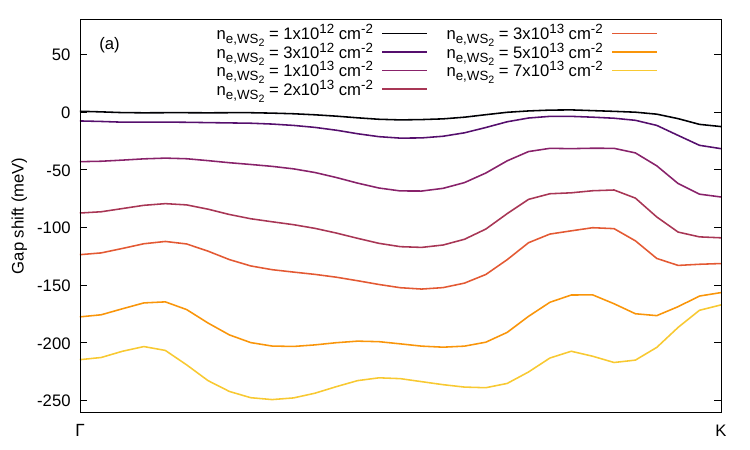}
		\includegraphics[width=.9\columnwidth]{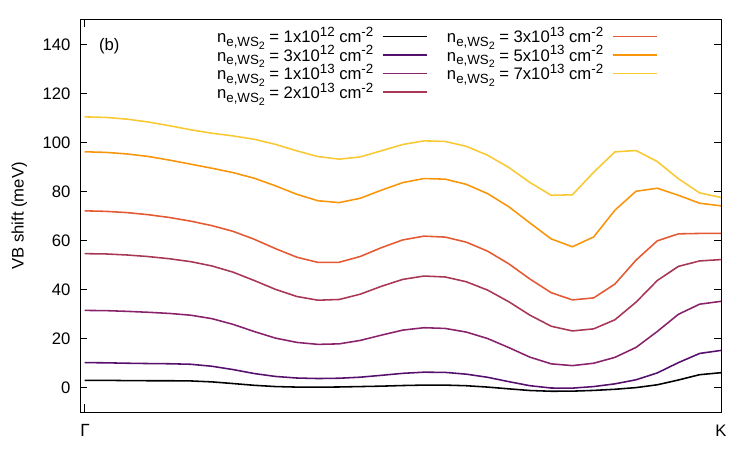}
		\caption{Calculated k-resolved transient band gap (a) and quasi-particle shifts of WS$_2$ VB (b) for different carrier densities for a carrier temperature of  $T=1000$\,K.} 
    \label{fig:shift_total}
    \end{figure}

     \newpage
    \clearpage
   
\subsection{Monolayer WS$_2$ results}

   \begin{figure}[h!t]
   	\centering
   	\includegraphics[width=\columnwidth]{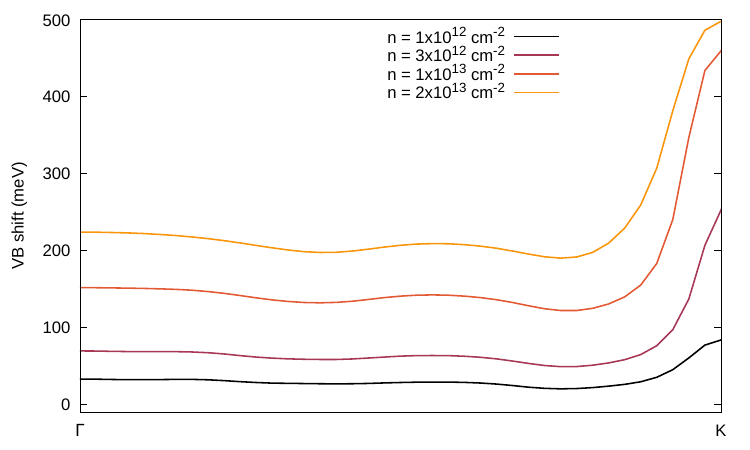}
   	\caption{Quasi-particle shifts of valence band (= minus hole energy) in monolayer WS$_2$ for different carrier densities at $T=1000$ K.} 
   	\label{fig:shift_monolayer}
   \end{figure}

   \newpage
   \clearpage
   
\subsection{Different contributions to valence-band renormalizations in photo\-excited heterostructure}

		\begin{figure}[h!t]
    \centering
    \includegraphics[width=\columnwidth]{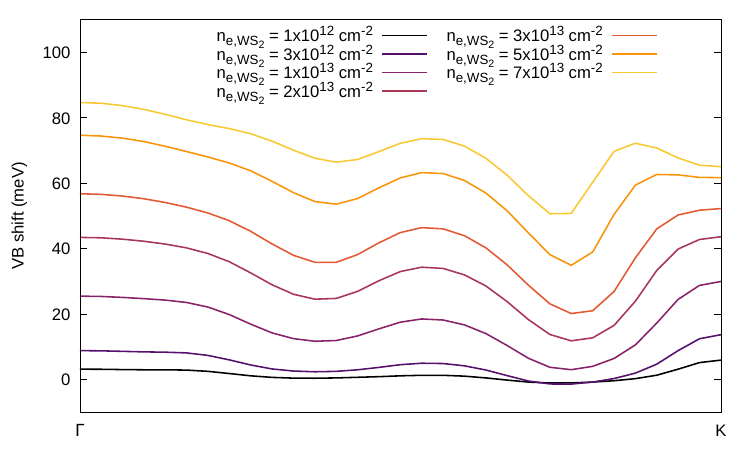}
    \caption{Only WS$_2$ GW contributions to valence-band shifts for different carrier densities at $T=1000$ K (sum of Fock and Montroll-Ward terms in Eq.~(\ref{eq:WS2_shift_diff})).} 
    \label{fig:shift_WS2_only}
    \end{figure}		
		
		\begin{figure}[h!t]
    \centering
    \includegraphics[width=\columnwidth]{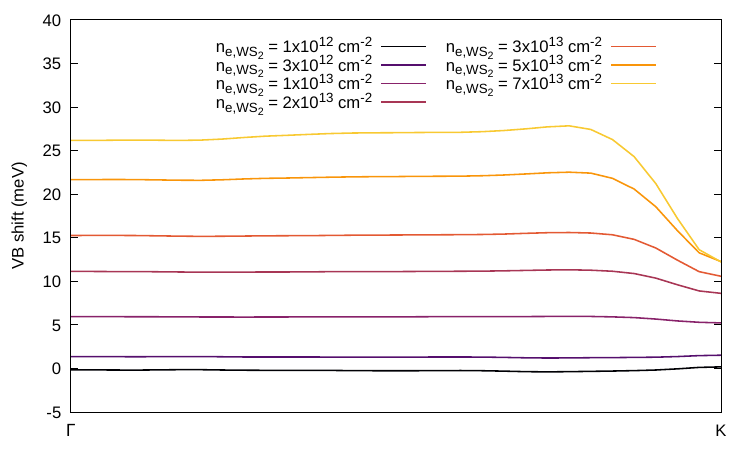}
    \caption{Only graphene GdW contributions to valence-band shifts for different carrier densities at $T=1000$ K (Eq.~(\ref{eq:Gdw_gr})).} 
    \label{fig:shift_graphene_only}
    \end{figure}
		
		\begin{figure}[h!t]
    \centering
    \includegraphics[width=\columnwidth]{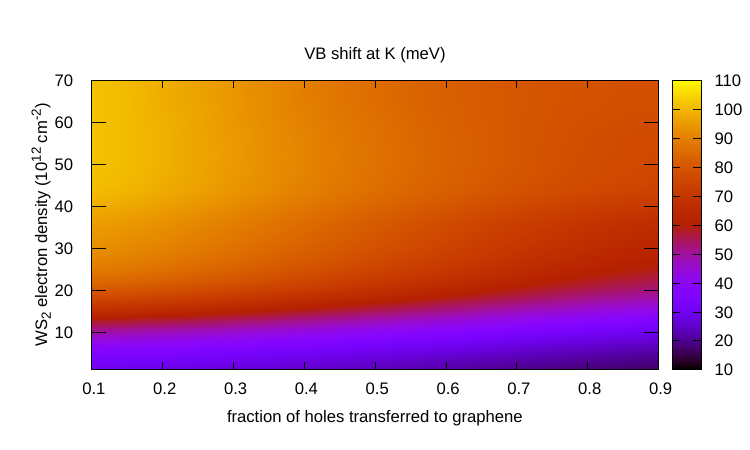}
    \caption{Dependence of WS$_2$ VB shift at K in meV including all above-mentioned contributions as a function of WS$_2$ electron density and fraction of holes transferred to graphene at $T=1000$ K.} 
    \label{fig:charge_transfer_scan}
    \end{figure}		
		
		\begin{figure}[h!t]
    \centering
    \includegraphics[width=\columnwidth]{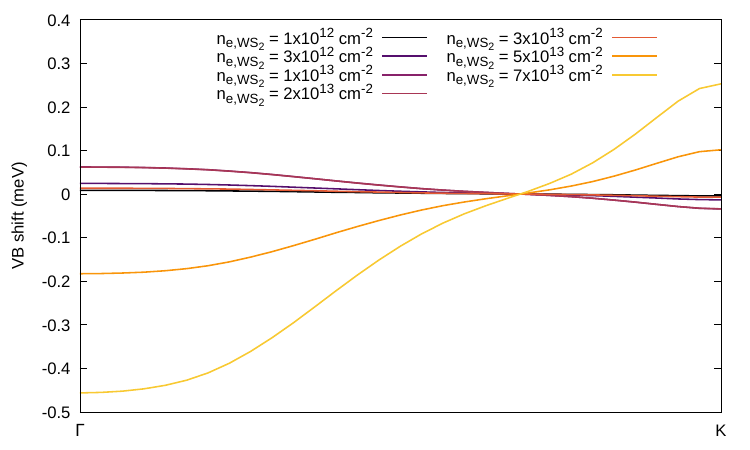}
    \caption{Only WS$_2$ Hartree contributions to valence-band shifts for different carrier densities at $T=1000$ K (Hartree term in Eq.~(\ref{eq:WS2_shift_diff})).} 
    \label{fig:shift_Hartree_only}
    \end{figure}
		
\clearpage

\bibliography{literature_merged}


\end{document}